\documentclass[a4paper]{article}
\usepackage{amsmath,amssymb}

\newcommand{\vp}{\mathbf{p}}
\newcommand{\vq}{\mathbf{q}}
\newcommand{\ip}{\int \frac{d^3 p}{(2 \pi)^3} \,}
\newcommand{\iq}{\int \frac{d^3 q}{(2 \pi)^3} \,}
\newcommand{\op}{\omega(\vp)}
\newcommand{\oq}{\omega(\vq)}
\newcommand{\osp}{\omega^2(\vp)}
\newcommand{\osq}{\omega^2(\vq)}

\newcommand{\app}{a(\vp)}

\newcommand{\adp}{a^\dag(\vp)}
\newcommand{\adq}{a^\dag(\vq)}
\newcommand{\xb}{\overline{x}}
\newcommand{\ix}{\int d^3_g x \,}
\newcommand{\my}{d^3_g y \,}
\newcommand{\mz}{d^3_g z \,}

\newcommand{\oj}{\omega(j)}
\newcommand{\ok}{\omega(k)}
\newcommand{\osj}{\omega^2(j)}

\newcommand{\inj}{\int d \mu(j) \,}
\newcommand{\ijo}{\int \frac{d\mu(j)}{\sqrt{2 \oj}} \,}
\newcommand{\mko}{\frac{d\mu(k)}{\sqrt{2 \ok}} \,}
\newcommand{\aj}{a(j)}

\newcommand{\adj}{a^\dag(j)}
\newcommand{\adk}{a^\dag(k)}
\newcommand{\pjx}{\phi_j(x)}
\newcommand{\psky}{\phi^\ast_k(y)}
\newcommand{\xt}{\mathbf{x}^\bot}
\newcommand{\yt}{\mathbf{y}^\bot}
\newcommand{\pt}{\mathbf{p}_\bot}
\newcommand{\qt}{\mathbf{q}_\bot}
\newcommand{\lip}{\int_0^\infty \frac{d p}{2 \pi} \,}
\newcommand{\liq}{\int_0^\infty \frac{d q}{2 \pi} \,}
\newcommand{\ipt}{\int \frac{d^2 p_\bot}{(2 \pi)^2} \,}
\newcommand{\iqt}{\int \frac{d^2 q_\bot}{(2 \pi)^2} \,}
\newcommand{\lop}{\omega(p,\pt)}
\newcommand{\loq}{\omega(q,\qt)}

\newcommand{\lap}{a(p,\pt)}

\newcommand{\ladp}{a^\dag(p,\pt)}
\newcommand{\ladq}{a^\dag(q,\qt)}
\begin{document}

  \date{}
  \title{\vskip-3em\textbf{ Relativistic Lee Model on Riemannian Manifolds  }}

  \author{Burak Tevfik Kaynak${\,}^{1}$\thanks{kaynakb@boun.edu.tr}\, and O. Teoman Turgut${\,}^{1,2}$\thanks{turgutte@boun.edu.tr}}

  \maketitle
  \begin{center}
    \vskip-3em 
    $^{1}\,$Department of Physics, Bogazici University, 34342 Bebek, Istanbul, Turkey \\
    $^{2}\,$Feza Gursey Institute, 34684 Kandilli, Istanbul, Turkey
  \end{center}
  \begin{abstract}
    We study the relativistic Lee model on static Riemannian manifolds. The model is constructed nonperturbatively through its resolvent, which is based on the so-called principal operator and the heat kernel techniques. It is shown that making the principal operator well-defined dictates how to renormalize the parameters of the model. The renormalization of the parameters  are the same in the light front coordinates as in the instant form. Moreover, the renormalization of the model on Riemannian manifolds agrees with the flat case. The asymptotic behavior of the renormalized principal operator in the large number of bosons limit implies that the ground state energy is positive. In $2+1$ dimensions, the model requires only a mass renormalization. We obtain rigorous bounds on the ground state energy for the $n$-particle sector of $2+1$ dimensional model.
  \end{abstract}
  \maketitle
  \section{Introduction}
    The Lee model is a simple field theory model, which requires a mass, coupling constant and wave function renormalization \cite{lee}. What is so special about the model is that the renormalizations can be carried out nonperturbatively. This is, therefore, a good testing ground for various new ideas and methods on interacting quantum field theories. In the original Lee model there are two fermion fields called $N$ and $V$, assumed to be so heavy that their energies are independent of the momentum, and a single relativistic real bosonic field named usually as $\theta$. The Lee model is amenable to exact analysis because there are two rather restricting conserved quantities. One of which is the conservation of the total number of the fermion species. Furthermore, the sum of bosons and $N$ type fermions is conserved. These highly constrain the theory allowing only a finite number of particles interacting at any given time. If we work with a complex scalar field, the situation changes drastically and the model becomes rather difficult \cite{wilson}. Although the renormalization is performed exactly, it is done so in a small number of particle sectors, and it is believed that the same prescriptions will continue to cure the divergences in all sectors. This is, of course, plausible since there are no other parameters in the theory. Once the physical $V$ particle is determined as a composite, the coupling constant is determined in such a way to make the scattering of $N$ and $\theta$ finite, thereby all the other physical processes should be well-defined (see a source theory approach to the model \cite{dittrich1}). The model is asymptotically free for $d<4$, and this point has been analyzed from the modern point of view in \cite{bender1,morris}. There is one subtle point, beyond a certain value of the renormalized coupling there appears a ghost state. This problem has been analyzed recently in the virtue of PT symmetry by Bender et al in \cite{bender2}. They show that by an appropriate redefinition of the norm, the ghost state can be turned into a physical state. Moreover in \cite{jones}, an equivalent hermitian Hamiltonian through a similarity transformation is constructed in the context of quasi-hermitian quantum mechanics. Even though the model is defined nonperturbatively, to understand the resulting spectrum remains largely as a challenge \cite{fuda}.

    The model is sufficiently rich, by restricting the total number of fermions to one, we can still get most of the interesting features. Moreover, one can assume that these fermions carry no momentum so that they have no recoil, and hence assumed to be fixed at the origin. This becomes equivalent to a two states system sitting at the origin interacting with relativistic real bosons \cite{marshall,bolsterli}. This is the version we will be working with so as to extend the model to Riemannian manifolds. The nonrelativistic version of this model, which is worked out beautifully in the book by Thirring and Henley \cite{thirring}, is still an interesting case study, yet in this case the coupling constant and wave function renormalizations are not needed. The study of the spectrum is still a nontrivial problem. To address these issues, there are some attempts in the literature \cite{north,nickle}. In \cite{rajeev}, while looking at some nonrelativistic problems which require nontrivial renormalizations, Rajeev introduced a new perspective. In this approach one attempts to renormalize the theory by working out the full resolvent in the Fock space of the system. The resolvent contains essentially all the information about the model. More interestingly, the bound states can be found through the zero eigenvalues of an operator, the so-called principal operator, which is parametrized by the energy in a nonlinear manner. Although one can write the resolvent, it is not possible to write down the quantum Hamiltonian of the renormalized theory. In the restricted Lee model, since the interaction is at a point, the renormalized model can be considered as a singular extension of the free bosonic Hamiltonian. This is analogous to the attractive delta function potential in two dimensions, which requires a coupling constant renormalization \cite{hoppe}. There one could also write down the resolvent but not the corresponding Hamiltonian. The interaction appears as a kind of boundary condition, this point of view originates from ideas of M. G. Krein on operators (see Albeverio and Kutasov \cite{albeverio} for a modern exposition). In \cite{rajeev}, this point of view is extended to the nonrelativistic Lee model, inspired from this work we develop the relativistic Lee model along the same lines. Having found the principal operator, and thus the resolvent, we can in principle work out all the physically important questions for all particle sectors.

    Following the heat kernel based methods,  developed in \cite{turgut}, we extend these ideas to the case of manifolds. The renormalizations are the same, of course, the resolvent contains information about the geometry through the heat kernel. The spectrum of the model is an interesting problem, we only attempt to partially understand it for large number of particles and show that the ground state energy remains positive in this limit. Developing new approximation methods for estimating the energy levels and scattering amplitudes remains as a challenge.

    The organization of the paper is as follows: In Section~(\ref{flat}), we will, first, construct the model in flat space-time through the approach, introduced in \cite{rajeev} without reviewing it. We show that the principal operator has a well-defined limit when the cut-off is removed, and the renormalized operator can be given by the renormalized mass and the renormalized coupling constant. Moreover, this limit determines the wave function renormalization constant. Afterwards, we specify how to impose the renormalization condition in this approach such that we convert the renormalized mass difference into the physical one by fixing the finite arbitrariness, which is left after renormalizing the parameters.

    In Section~(\ref{manifold}), we will apply the ideas, presented in \cite{turgut}, to the relativistic Lee model on a general static Riemannian manifolds. It is shown that the regularization of the ultra-violet divergence in the theory can be established through the short-time expansion of the heat kernel if the point interaction is introduced by a convolution of the bosonic field with a heat kernel. It is found that the divergence structure of the model in the manifold case is exactly the same as the flat case.

    In Section~(\ref{als}), we study the asymptotic behaviors of the renormalized principal operator in the large number of bosons limit in both
    flat and manifold cases. In this limit, it is shown that the leading behavior of the theory changes substantially. The ultra-static space-time $\mathbb{R} \times \mathbb{H}^3$ is given as an example.

    In Section~(\ref{2p1}), we study the model in $2+1$ dimensions. The advantage is that it is simple and requires only a mass renormalization. This allows us to find rigorous bounds on the ground state energy for $n$-particle sector, thus illustrating the power of this method.

    In Appendix, the same techniques are tested in an oblique light-front coordinates.
  \section{Relativistic Lee Model in $\mathbb{R}^{3+1}$}\label{flat}
    The model which we will construct in this section describes the interaction between a field of relativistic bosons and a heavy fermionic source with an internal degree of freedom which actually corresponds to \textit{two distinct states} of the source. Since the source is heavy, we can effectively consider it as sitting at some fixed point in space-time. This results in the neglect of recoil for the source, which means that the energies of the states do not depend on their momentum \cite{schweber}. The cut-off Hamiltonian of the model in matrix form is given by
    \begin{equation}\label{1}
      H_\epsilon = H_0 \left[ \chi_+ \otimes \chi_+^\dag + Z(\epsilon) \chi_- \otimes \chi_-^\dag \right] + H_{I,\epsilon} \,,
    \end{equation}
    where $H_0$ and $H_{I,\epsilon}$ are the free and the interaction parts of the cut-off Hamiltonian respectively and are given by
    \begin{align}
    H_0 &= \ip \op \adp \app \,,\\
    H_{I,\epsilon} &= Z(\epsilon) \mu(\epsilon) \frac{1 - \sigma_3}{2} + \sqrt{Z(\epsilon)} \lambda (\epsilon) \left[ \sigma_+ \phi^{(-)}_\epsilon(0) + \sigma_-
    \phi^{(+)}_\epsilon(0) \right] \,.
    \end{align}
    At this moment $\epsilon$ is an unspecified cut-off prescription, the meaning of which will become clear when we will renormalize the parameters of the theory. Here, $\phi^{(\pm)}_\epsilon(0)$ are the positive and negative frequency parts of the real bosonic field, respectively, defined through this cut-off prescription. A more precise definiton of these field operators will be given in the manifold case. $\chi_\pm$ in Eq.~(\ref{1}) are the standard spin states, which describe the two states of the system. Due to the fact that there can be various divergences hidden inside the theory, we allow two states of the system to have different normalizations. With hindsight we choose this to be $\chi_-$ state.

    The theory has a conserved charge, which can be written as,
    \begin{equation}
      Q = - \frac{1 - \sigma_3}{2} + \ip \adp \app \,.
    \end{equation}
    This means that the theory decouples into independent sectors as $\mathcal{F}_\mathcal{B}^{(n)} \otimes \chi_+ \oplus \mathcal{F}_\mathcal{B}^{(n-1)} \otimes \chi_-$.

    The construction of the model is merely based on finding the resolvent of the cut-off Hamiltonian, which describes the system completely. While computing the resolvent, one introduces the principal operator $\Phi(E)$, that can be regarded as an effective Hamiltonian of the theory. The reason that the Krein formula can be applied lies in the observation that there is a constraining conserved quantity, namely, $Q$. The idea of using this operator comes from the fact that the zero eigenvalues of it determine implicitly the bound state energies of the theory. The ultra-violet divergence takes place in the theory when the size of the source goes to zero, which causes the difference of the energy levels to become infinite. In the following, by renormalization it is solely meant to search for a well-defined limit of the cut-off Hamiltonian in matrix form as $\epsilon \rightarrow 0^+$. This is accomplished by curing the principal operator in the same limit.

    Following Rajeev \cite{rajeev} the cut-off Hamiltonian minus energy is given in a $2\times2$ block form,
    \begin{equation} \label{ch}
      H_\epsilon -E = \begin{bmatrix}
                        H_0 - E & \ \sqrt{Z(\epsilon)} \lambda(\epsilon) \phi^{(-)}_\epsilon(0)  \\
                        \sqrt{Z(\epsilon)} \lambda(\epsilon) \phi^{(+)}_\epsilon(0) & \ \ Z(\epsilon) \left[ H_0 - E + \mu(\epsilon) \right] \\
                      \end{bmatrix} \,.
    \end{equation}
    The resolvent is simply the formal inverse of Eq.~(\ref{ch}) and this inverse can be calculated algebraically. If the Hamiltonian is parametrized as,
    \begin{equation}
      H_\epsilon - E = \begin{pmatrix}
                         a & b^\dag \\
                         b & d \\
                       \end{pmatrix} \,,
    \end{equation}
    and if the resolvent is parametrized as,
    \begin{equation}
      R_\epsilon (E) = \begin{pmatrix}
                         \alpha & \beta^\dag \\
                         \beta & \delta
                       \end{pmatrix} \,,
    \end{equation}
    then one ends up with the following algebraic equalities, which allow one to calculate the resolvent,
    \begin{align}
      \alpha &= a^{-1} + a^{-1} b^\dag \left( d -b a^{-1}  b^\dag \right)^{-1} b a^{-1}  \,,\\
      \beta &= - \left( d -b a^{-1}  b^\dag \right)^{-1} b a^{-1} \,, \label{beta}\\
      \delta &= \left( d -b a^{-1}  b^\dag \right)^{-1} = \delta^\dag \,, \label{delta}\\
      \Phi &= d -b a^{-1}  b^\dag \,. \label{p}
    \end{align}
    Equation~(\ref{p}) is just the cut-off principal operator and is given by
    \begin{equation}
      \Phi_\epsilon(E) =  Z(\epsilon) \left\{ H_0 - E + \mu(\epsilon) - \lambda^2(\epsilon) \ip \iq \frac{\app}{\sqrt{2 \op}} \frac{1}{(H_0 - E)} \frac{\adq}{\sqrt{2 \oq}} \right\} \,,
    \end{equation}
    where $E$ is considered as a complex parameter and the formulae below should be analytically continued to their largest domains of analyticity. As one can easily notice the annihilation and the creation operators are in the wrong order with respect to normal ordering prescription, so we should normal-order them. After being normal-ordered, the principal operator becomes
    \begin{align}\label{phi1}
      \Phi_\epsilon(E) &= Z(\epsilon) \Bigg\{ H_0 - E + \mu(\epsilon) - \lambda^2(\epsilon) \ip \frac{1}{2 \op} \frac{1}{H_0 - E + \op} \nonumber \\
      & \qquad - \lambda^2(\epsilon) \ip \iq  \frac{\adq}{\sqrt{2 \oq}} \frac{1}{H_0 - E + \oq + \op} \frac{\app}{\sqrt{2 \op}} \Bigg\} \,.
    \end{align}
    The two fractions in the fourth term above can be united by a Feynman parametrization, and then can be exponentiated as
    \begin{align}
      \frac{1}{\op} \frac{1}{H_0 - E + \op} &= \int_0^1 d \xi \, \frac{1}{\left[ (H_0 - E) \xi + \op \right]^2} \nonumber \\
      &= \int_0^1 d \xi \, \int_0^\infty ds \, e^{- s \op} e^{-s (H_0 - E) \xi} \nonumber \\
      &= \int_0^\infty ds \, e^{- s \op} \frac{1}{H_0 -E} \left[ 1 - e^{- s (H_0 - E)} \right] \,.
    \end{align}
    In order to evaluate the momentum integral one more identity is needed, and this is the so-called subordination identity:
    \begin{equation}
      e^{-s \op} = \frac{s}{2\sqrt{\pi}} \int_0^\infty du \, \frac{1}{u^{3/2}} e^{-s^2/4u} e^{-u \osp} \,.
    \end{equation}
    With the help of that identity, we can convert $\op$ in the exponential into ${\osp}$ such that after calculating the momentum integral the second term can be given by
    \begin{align} \label{se}
      \ip \frac{1}{2 \op} \frac{1}{H_0 - E + \op} &= \frac{1}{4\sqrt{\pi}} \int_\epsilon^\infty du \, \frac{e^{- u m^2}}{u^{3/2}} \ip e^{- u p^2} \int_0^\infty ds \, s e^{-s^2/4u} \frac{1}{H_0 -E} \left[ 1 - e^{- s (H_0 - E)} \right] \nonumber \\
      &= \frac{1}{32 \pi^2} \int_\epsilon^\infty du \, \frac{e^{- u m^2}}{u^{3/2}} \int_0^\infty ds \, s e^{-s^2/4} \frac{\left[ 1 - e^{- s \sqrt{u} (H_0 - E)} \right]}{\sqrt{u} (H_0 -E)} \,.
    \end{align}
    The use of that identity is to convert the divergence of the momentum integral into a divergence emerging from the lower limit of the $u$-integral, which is now mollified by $\epsilon$, explicitly. The momentum integral is no longer divergent, and thus, can safely be computed. If the same calculations are done step by step for the three fractions in the fifth term in Eq.~(\ref{phi1}) without calculating the momentum integral, one obtains
    \begin{align}\label{no}
      &\frac{1}{\sqrt{2 \oq}} \frac{1}{H_0 - E + \oq + \op} \frac{1}{\sqrt{2 \op}}
      = \frac{2}{\pi} \int_0^\infty d \alpha \, \int_0^\infty d \beta \, \int_0^\infty ds \, e^{- \oq (s + \alpha^2)} e^{- \op (s + \beta^2)} e^{-s (H_0 -E)} \nonumber \\
      &= \frac{1}{2\pi^2} \int_0^\infty ds \, \int_0^\infty d \alpha \, \left( s + \alpha^2 \right)\int_0^\infty d \beta \, \left(s + \beta^2\right)
      \int_0^\infty du_1 \, \frac{e^{- (s + \alpha^2)^2/4 u_1}}{u_1^{3/2}}  \int_0^\infty du_2 \, \frac{e^{- (s + \beta^2)^2/4 u_2}}{u_2^{3/2}} \nonumber \\
      & \qquad \times e^{- u_1 \osq} e^{- u_2 \osp} e^{-s (H_0 -E)} \,.
    \end{align}
    After plugging Eqs.~(\ref{se}) and (\ref{no}) into Eq.~(\ref{phi1}), the cut-off principal operator becomes
    \begin{align}
      \Phi_\epsilon(E) &= Z(\epsilon) \Bigg\{ H_0 - E + \mu(\epsilon) - \frac{\lambda^2(\epsilon)}{32 \pi^2} \int_\epsilon^\infty du \, \frac{e^{- u m^2}}{u^{3/2}} \int_0^\infty ds \, s e^{-s^2/4} \frac{\left[ 1 - e^{- s \sqrt{u} (H_0 - E)} \right]}{\sqrt{u} (H_0 -E)} \nonumber \\
      & \qquad - \frac{\lambda^2(\epsilon)}{2 \pi^2} \int_0^\infty ds \, \int_0^\infty d \alpha \, \left( s + \alpha^2 \right)\int_0^\infty d \beta \, \left(s + \beta^2\right) \int_0^\infty du_1 \, \frac{e^{- (s + \alpha^2)^2/4 u_1}}{u_1^{3/2}} \int_0^\infty du_2 \, \frac{e^{- (s + \beta^2)^2/4 u_2}}{u_2^{3/2}} \nonumber \\
      & \qquad \times \ip \iq e^{- u_1 \osq} e^{- u_2 \osp} \adq e^{-s (H_0 -E)} \app \Bigg\} \,.
    \end{align}

    It is now time to renormalize the cut-off principal operator. If the exponential in the second integral term is expanded in power series in $s$, one can notice that the only terms which produce divergence are just the ones which are up to order $s^2$. On the basis of this expansion, one can redefine the coupling constant and the mass, whereby the cut-off principal operator can, easily, be regularized. Thus, we are able to achieve the renormalized counterparts of both those parameters and the principal operator. In order to accomplish this, it is appropriate to divide the principal operator by the square of the coupling constant. The main difference of the relativistic Lee model from the nonrelativistic one resides in not only that there is a coupling constant renormalization besides the mass renormalization but also there is a wave function renormalization. Therefore, the ratio of the principal operator to the square of the coupling constant, $\Phi(E)/\lambda^2$, should be renormalized instead of just the principal operator, $\Phi(E)$, in the relativistic Lee model. In the light of the above discussion, one can renormalize whole the parameters of the model. The following choices regularize the principal operator by canceling the divergences,
    \begin{align}
      \frac{\mu(\epsilon)}{\lambda^2(\epsilon)} &= \frac{\mu_R}{\lambda_R^2} + \frac{1}{32 \pi^2} \int_\epsilon^\infty du \, \frac{e^{- u m^2}}{u^{3/2}} \int_0^\infty ds \, s^2 e^{-s^2/4} \,, \label{ml}\\
      \frac{1}{\lambda^2(\epsilon)} &= \frac{1}{\lambda_R^2} - \frac{1}{64 \pi^2} \int_\epsilon^\infty du \, \frac{e^{- u m^2}}{u} \int_0^\infty ds \, s^3 e^{-s^2/4} \,, \label{l2}
    \end{align}
    and then the principal operator is given by
    \begin{align}
      \frac{\Phi_\epsilon(E)}{\lambda^2(\epsilon)} &= Z(\epsilon) \Bigg\{ \frac{(H_0 - E)}{\lambda^2_R} + \frac{\mu_R}{\lambda^2_R} - \frac{1}{32 \pi^2} \int_\epsilon^\infty du \, \frac{e^{- u m^2}}{u^{3/2}} \int_0^\infty ds \, s e^{-s^2/4} \frac{1}{\sqrt{u} (H_0 -E)} \nonumber \\
      & \qquad \times \left[ 1 - s \sqrt{u} (H_0 - E) + \frac{1}{2} s^2 u (H_0 - E)^2 - e^{- s \sqrt{u} (H_0 - E)} \right] \nonumber \\
      & \qquad - \frac{1}{2 \pi^2} \int_0^\infty ds \, \int_0^\infty d \alpha \, \left( s + \alpha^2 \right)\int_0^\infty d \beta \, \left(s + \beta^2\right) \int_0^\infty du_1 \, \frac{e^{- (s + \alpha^2)^2/4 u_1}}{u_1^{3/2}} \int_0^\infty du_2 \, \frac{e^{- (s + \beta^2)^2/4 u_2}}{u_2^{3/2}} \nonumber \\
      & \qquad \times \ip \iq e^{- u_1 \osq} e^{- u_2 \osp} \adq e^{-s (H_0 -E)} \app \Bigg\} \,.
    \end{align}
    We notice that the subtractions in the second line resemble the regularization of the infinite Fredholm determinants. This is analogous to the quantum effective action calculations via regularized determinants in the path integral formalism. It is obvious that this operator has a well-defined limit as $\epsilon \rightarrow 0^+$ when both sides are divided by $Z(\epsilon)$:
    \begin{equation}
      \lim_{\epsilon \rightarrow 0^+} \frac{\Phi_\epsilon (E)}{\lambda^2(\epsilon) Z(\epsilon)} = \frac{\Phi_R (E)}{\lambda^2_R} \,,
    \end{equation}
    and the renormalized principal operator in terms of the renormalized mass and the renormalized coupling constant can be given by
    \begin{align}
      \frac{\Phi_R(E)}{\lambda_R^2} &= \frac{(H_0 - E)}{\lambda^2_R} + \frac{\mu_R}{\lambda^2_R} - \frac{1}{32 \pi^2} \int_0^\infty du \, \frac{e^{- u m^2}}{u^{3/2}} \int_0^\infty ds \, s e^{-s^2/4} \frac{1}{\sqrt{u} (H_0 -E)} \nonumber \\
      & \qquad \times \left[ 1 - s \sqrt{u} (H_0 - E) + \frac{1}{2} s^2 u (H_0 - E)^2 - e^{- s \sqrt{u} (H_0 - E)} \right] \nonumber \\
      & \qquad - \frac{1}{2 \pi^2} \int_0^\infty ds \, \int_0^\infty d \alpha \, \left( s + \alpha^2 \right)\int_0^\infty d \beta \, \left(s + \beta^2\right) \int_0^\infty du_1 \, \frac{e^{- (s + \alpha^2)^2/4 u_1}}{u_1^{3/2}} \int_0^\infty du_2 \, \frac{e^{- (s + \beta^2)^2/4 u_2}}{u_2^{3/2}} \nonumber \\
      & \qquad \times \ip \iq e^{- u_1 \osq} e^{- u_2 \osp} \adq e^{-s (H_0 -E)} \app \,.
    \end{align}
    It is easily seen that this limit also fixes the wave function renormalization constant $Z(\epsilon)$ to be equal to $\lambda_R^2/\lambda^2(\epsilon)$. Now the elements of the resolvent can also be given in terms of the renormalized principal operator. To see this, we look at the expression for $\alpha$,
    \begin{equation}
      \alpha = \frac{1}{H_0 - E} + \frac{1}{H_0 - E} \phi_\epsilon^{(-)}(0) \frac{Z(\epsilon) \lambda^2(\epsilon)}{\Phi_\epsilon(E)} \phi_\epsilon^{(+)}(0) \frac{1}{H_0 - E} \,. \\
    \end{equation}
    If we now use the wave function renormalization constant in this expression, we may take the limit $\epsilon \rightarrow 0^+$, giving us
    \begin{equation}
      \alpha = \frac{1}{H_0 - E} + \frac{1}{H_0 - E} \phi^{(-)}(0) \frac{\lambda^2_R}{\Phi_R(E)} \phi^{(+)}(0) \frac{1}{H_0 - E} \,.
    \end{equation}
    Similarly, for the others, we find
    \begin{align}
      \beta &= - \frac{\lambda_R}{\Phi_R(E)} \phi^{(-)}(0) \frac{1}{H_0 - E} \,, \\
      \delta &= \frac{1}{\Phi_R(E)} \,.
    \end{align}
    These equations tell us that zero eigenvalues of the renormalized principal operator determine the bound states and the corresponding energies as nonlinear eigenvalue equations. Notice that the renormalized operator $\Phi_R(E)$ converts a divergent linear problem in the Schr\"{o}dinger picture into a highly nonlinear but a well-defined problem.

    It is also important to know how the divergences are controlled by the cut-off parameter $\epsilon$ in the redefinition of the mass and the coupling constant. In order to find it out, the integrals in Eqs.~(\ref{ml}) and (\ref{l2}) should be calculated. The cut-off dependent parameters can be given as an asymptotic series in $\epsilon$,
    \begin{align}
      \frac{\mu(\epsilon)}{\lambda^2(\epsilon)} \simeq \frac{\mu_R}{\lambda_R^2} + \frac{1}{8 \pi^{3/2}} \frac{1}{\sqrt{\epsilon}} \quad \mathrm{as} \quad \epsilon \rightarrow 0^+ \,, \qquad
      \frac{1}{\lambda^2(\epsilon)} \simeq \frac{1}{\lambda_R^2} + \frac{1}{8 \pi^2} \ln \epsilon \quad \mathrm{as} \quad \epsilon \rightarrow 0^+ \,.
    \end{align}

    To make contact with the usual perturbative renormalization, we will recast the Hamiltonian into a renormalized part and a counterterm Hamiltonian. We will see that there are no other counterterms needed other than the ones existing already in the original Hamiltonian. So as to establish that, we should go back in the calculations and replace $Z(\epsilon)$ by $\lambda_R^2/\lambda^2(\epsilon)$. The mass term of the source becomes
    \begin{align}
      Z(\epsilon) \mu(\epsilon) \frac{1 - \sigma_3}{2} &= \frac{\mu(\epsilon)}{\lambda^2(\epsilon)} \lambda^2_R \frac{1 - \sigma_3}{2} \nonumber \\
      &= \left( \mu_R + \lambda^2_R \Delta \mu \right) \frac{1 - \sigma_3}{2} \,,
    \end{align}
    in which the term $\Delta\mu$ is nothing but the divergent part in the redefinition of the mass. The same replacement should also be done in the interaction terms and one can get
    \begin{align}
      \sqrt{Z(\epsilon)} \lambda (\epsilon) \left[ \sigma_+ \phi^{(-)}_\epsilon(0) + \sigma_- \phi^{(+)}_\epsilon(0) \right] &= \lambda_R \left[ \sigma_+ \phi^{(-)}(0) + \sigma_- \phi^{(+)}(0) \right] \,.
    \end{align}
    The next step to perform is plug those into the Hamiltonian such that the renormalized Hamiltonian can be determined. After plugging them, the Hamiltonian becomes,
    \begin{align}
      H_\epsilon &= H_0 \left[ \chi_+ \otimes \chi_+^\dag + Z(\epsilon) \chi_- \otimes \chi_-^\dag \right] + Z(\epsilon) \mu(\epsilon) \frac{1 - \sigma_3}{2} + \sqrt{Z(\epsilon)}\lambda (\epsilon) \left[ \sigma_+ \phi^{(-)}_\epsilon(0) + \sigma_-
      \phi^{(+)}_\epsilon(0) \right] \nonumber \\
      &= H_0 \left[ \chi_+ \otimes \chi_+^\dag + Z(\epsilon) \chi_- \otimes \chi_-^\dag \right] + \left(\mu_R + \lambda^2_R \Delta \mu\right) \frac{1 - \sigma_3}{2} + \lambda_R \left[ \sigma_+ \phi^{(-)}(0) + \sigma_- \phi^{(+)}(0) \right] \,.
    \end{align}
    We know from the theory of renormalization that if one would like to give the Hamiltonian of the theory in terms of renormalized parameters instead of bare or cut-off parameters, then the bare Hamiltonian is given by the renormalized Hamiltonian containing only the renormalized parameters plus the appropriate counterterms. Therefore if we choose the cut-off Hamiltonian as
    \begin{equation}
      H_\epsilon = H_R + H_0 [Z(\epsilon) - 1] \chi_- \otimes \chi_-^\dag + \lambda^2_R \Delta \mu \frac{1 - \sigma_3}{2} \,,
    \end{equation}
    then the renormalized Hamiltonian of the theory can be given by
    \begin{equation}
      H_R = H_0 \left[ \chi_+ \otimes \chi_+^\dag + \chi_- \otimes \chi_-^\dag \right] + \mu_R \frac{1 - \sigma_3}{2} + \lambda_R \left[ \sigma_+ \phi^{(-)}(0) + \sigma_- \phi^{(+)}(0) \right] \,.
    \end{equation}
    The renormalized Hamiltonian $H_R$ should not be confused with what we call the quantum Hamiltonian $H_Q$, which determines the time evolution of the quantum system. The resolvent that we have found in the Fock space should correspond to the resolvent of the Hamiltonian $H_Q$ defined in this Fock space. The existence of this Hamiltonian cannot be proved by a straightforward application of the resolvent convergence as is done for a different model in \cite{dimock}. This question is delicate in our case. However its resolvent can be explicitly derived, this Hamiltonian may not be written as an explicit formula.

    Although the renormalized parameters had been found, we did not complete the renormalization. Since after regularizing the parameters by removing the divergences, there remains a finite arbitrariness \cite{itzykson}. In order to fix these finite parts, which results in determining the physical parameters of the theory, one should impose the renormalization conditions. In perturbative field theories, these conditions should be imposed on the superficially divergent Green's functions to determine the coefficients of the counterterms and one demands that Green's functions satisfy them order by order if these conditions are satisfied to lowest order. In our formulation we should also specify similar conditions. In this approach the Schr\"{o}dinger equation is replaced by the equation $\Phi(E)\Psi=0$. So a natural choice is the one related to the simple composite which consists of a single boson and $\chi_+$ state giving us a dressed $\chi_-$ state. We can fix the mass difference of $\chi_-$ and $\chi_+$. Therefore, we impose the following,
    \begin{equation}
      \Phi_R(E=\mu_p)|0\rangle \equiv 0 \,,
    \end{equation}
    where $\mu_p$ is the physical mass difference. If the calculations of the principal operator are followed backwards, one can obtain a much more compact version of the principal operator. After a little algebra, we get
    \begin{align}
      \Phi_R ( E ) &= H_0 - E + \mu_R - \frac{\lambda_R^2}{2} \ip \,\left[ \frac{1}{\op} \frac{1}{H_0 - E + \op} - \frac{1}{\osp} + \frac{H_0 - E}{\omega^3(\vp)} \right] + \cdots \,,
    \end{align}
    where the dots stand for the normal-ordered interaction term. Incidentally, if we could expand the first term in the integral into a power series in $H_0 - E$, the second and the third terms are canceled, leading to a series of finite terms. If we add and subtract the second line above with $E = \mu_p$ and $H_0 = 0$, the resultant operator will satisfy the desired condition and fix the finite part of the renormalization.
    \begin{align}
      \Phi_R ( E ) &= H_0 - E + \mu_p + \frac{\lambda_R^2}{2} \ip \left[ \frac{1}{\op} \frac{1}{ - \mu_p + \op} - \frac{1}{\osp} + \frac{- \mu_p}{\omega^3(\vp)} \right] \nonumber \\
      & \qquad - \frac{\lambda_R^2}{2} \ip \left[ \frac{1}{\op} \frac{1}{H_0 - E + \op} - \frac{1}{\osp} + \frac{H_0 - E}{\omega^3(\vp)} \right] \cdots \nonumber \\
      \Phi_R ( E ) &= \left( H_0 - E + \mu_p \right) \left\{ 1 + \frac{\lambda_R^2}{2} \ip \left[ \frac{1}{\op \left[H_0 - E + \op\right] \left[ -\mu_p + \op \right]} - \frac{1}{\omega^3(\vp)} \right] \right\} \label{phi} \nonumber \\
      & \qquad - \lambda_R^2 \ip \iq  \frac{\adq}{\sqrt{2 \oq}} \frac{1}{H_0 - E + \oq + \op} \frac{\app}{\sqrt{2 \op}} \,.
    \end{align}
    In Section (\ref{als}) we will use Eq.~(\ref{phi}) to analyze the asymptotic limit of the theory with the assistance of the asymptotic limit of the principal operator.
  \section{Relativistic Lee Model on Riemannian Manifolds}\label{manifold}
    First, we will summarize the necessary tools before going into the details of the construction of the model on Riemannian manifolds \cite{fulling}. We consider $4d$ Riemannian manifold equipped with a metric structure which is static. That is to say, there is a timelike Killing vector field and there is a family of spacelike hypersurfaces orthogonal to the Killing vector everywhere. Alternatively, there is a coordinate system in which not only are the metric components $g_{\mu\nu}$ independent of the time coordinate, but also $g_{0j}=0$ for $j\neq0$.

    It is assumed that the action of a bosonic field can be given by
    \begin{equation}
      S = \int d^{4} x \, \sqrt{|g|} \frac{1}{2} \left( g^{\mu \nu} \partial_\mu \phi \partial_\nu \phi - m^2 \phi^2 - \xi R \phi^2 \right) \,,
    \end{equation}
    where $\xi$ is a dimensionless constant and $R$ is the curvature scalar of the manifold. The indefinite analogue of the Laplace-Beltrami operator, the so-called wave operator obtained through the covariant derivative, and the resulting field equations are given by, respectively,
    \begin{align}
      & \square \phi = \frac{1}{\sqrt{g}} \partial_\mu \left[ g^{\mu\nu} \sqrt{g} \partial_\nu \phi \right] \,, \\
      & \square \phi + (m^2 + \xi R) \phi = 0 \,.
    \end{align}
    Since the metric is static, the field equations can be solved by separation of variables, which results in the eigenvalue equation for the operator $L$,
    \begin{align}
      L\phi &= g_{00} \left[ \frac{1}{\sqrt{|g|}} \partial_j ( \sqrt{|g|} g^{jk} \partial_k \phi) + (m^2 + \xi R) \phi \right] \,, \label{l} \\
      L \phi_j &= \osj \phi_j \,,
    \end{align}
    where the bosonic field decomposed as
    \begin{equation}
      \phi(t,x) = \phi_j (x) e^{\mp i\oj t} \,.
    \end{equation}
    The operator $L$ is formally self-adjoint with respect to $\mathcal{L}_p^2$ inner product defined through
    \begin{equation}\label{ip}
      \left( \phi_1, \phi_2 \right) = \int d^3 x \, \sqrt{|g|} g^{00} \phi^\ast_1 (x) \phi_2 (x) \,.
    \end{equation}
    Any function in the Hilbert space defined by that inner product can be expanded in terms of the solutions of Eq.~(\ref{l}), namely the eigenfunctions of that operator, as
    \begin{equation}
      \phi (x) = \inj \phi(j) \pjx \,,
    \end{equation}
    where $\inj$ is the measure and it contains point spectrum or discrete spectrum or both. By means of that expansion, the scalar product defined by Eq.~(\ref{ip}) can be given by
    \begin{equation}
      \left( \phi_1, \phi_2 \right) = \inj \phi^\ast_1 (j) \phi_2(j) \,.
    \end{equation}
    We have also the orthonormality and the completeness relations. They are,
    \begin{align}
      \delta (j,k) &= \int d^3 x \, \sqrt{|g|} g^{00} \phi^\ast_j(x) \phi_k(x) \,, \\
      \delta_g^{(3)}(x,x') &= \int d\mu(j) \, \phi^\ast_j(x) \phi_j(x') \,.
    \end{align}
    The general solution of the field equation can be decomposed into positive and negative parts and they are given by, respectively,
    \begin{align}
      \phi(t,x) &= \ijo \left[ \aj \pjx e^{-i \oj t} + \adj \phi^\ast_j(x) e^{i \oj t} \right] \,, \\
      \phi^{(+)}(x) &= \ijo \pjx \aj  \,, \\
      \phi^{(-)}(x) &= \ijo \phi^\ast_j(x) \adj \,,
    \end{align}
    where $\aj$ and $\adj$ are the annihilation and the creation operators.
    A conjugate momentum and a Hamiltonian should be defined in order to quantize the field canonically. The conjugate momentum is
    \begin{equation}\label{cm}
      \pi(t,x) = g^{00} \sqrt{|g|} \partial_0 \phi \,,
    \end{equation}
    and the Hamiltonian is just the Legendre transform of the Lagrangian. With the help of them, one can calculate the equal-time canonical commutation relations both between the field and the conjugate momentum, and then between the creation and annihilation operators.
    \begin{equation}
      \left[ \phi(t,x) , \pi(t,x') \right] = i \sqrt{|g|} g^{00} \delta_g^{(3)} (x,x')\,, \quad \left[ \aj , \adk \right] = \delta(j,k) \,.
    \end{equation}
    The free Hamiltonian in terms of creation and annihilation operators is given by
    \begin{equation}
      H_0 = \inj \oj \adj \aj \,.
    \end{equation}
    Since the source is heavy and essentially sits at a point in space, one has to find a way to describe this situation. We use the same trick which was used in \cite{turgut}. The interaction is introduced by a convolution of the bosonic field with a heat kernel whose index is just a short-time cut-off. In the limit as the cut-off goes to zero, the heat kernel becomes a Dirac delta function and hence the convolution in this limit allows us to find the interaction occurring at some fixed point in space. Utilizing the short-time behavior of the heat kernel is a nice way to analyze and control the high energy behavior of the expressions, so this allows us to deal with the ultra-violet divergence in the theory. The cut-off Hamiltonian of the theory on Riemannian manifold, specified previously, is
    \begin{align}
      H_\epsilon &= H_0 \left[ \chi_+ \otimes \chi_+^\dag + Z(\epsilon) \chi_- \otimes \chi_-^\dag \right] + Z(\epsilon) \mu(\epsilon) \frac{1 - \sigma_3}{2} + \sqrt{Z(\epsilon)} \lambda (\epsilon) \left[ \sigma_+ \phi^{(-)}_\epsilon(\xb) + \sigma_-
      \phi^{(+)}_\epsilon(\xb) \right] \,,
    \end{align}
    in which the smeared out positive and negative frequency parts of the field are given by
    \begin{align}
      \phi^{(+)}_\epsilon(\xb) &= \ix K_{\epsilon/2}(\xb,x) \phi^{(+)}(x) \,,\\
      \phi^{(-)}_\epsilon(\xb) &= \ix K_{\epsilon/2}(\xb,x) \phi^{(-)}(x) \,,
    \end{align}
    where $\int d^3_g x \equiv \int d^3 x \, \sqrt{|g(x)|} g^{00}(x)$, $\overline{x}$ is a fixed point on the manifold, and $\epsilon/2$ is chosen for convenience.

    Before carrying on, we would like to list the important properties of the heat kernel, \cite{rosenberg}, which we will use throughout this and the next section:
    \begin{align}
      K_u(x,y) &= K_u(y,x) \,, & \text{Symmetry property} \,, \nonumber \\
      L K_u(x,y) &= \frac{\partial}{\partial u} K_u(x,y) \,, & \text{Heat equation} \,, \nonumber \\
      \lim_{u \rightarrow 0^+} K_u(x,y) &= \delta_g^{(3)}(x,y) \,, & \text{Initial condition} \,, \nonumber \\
      \int_\mathcal{M} d_g^3 z \, K_{u_1}(x,z) K_{u_2}(z,y) &= K_{u_1 + u_2} (x,y) \,, & \text{Reproducing property} \,, \nonumber \\
      K_u(x,y) &\geq 0 \quad \text{for all } u \,, & \text{Positivity} \,.
    \end{align}

    The resolvent is again the formal inverse of the operator,
    \begin{equation}
      H_\epsilon -E = \begin{bmatrix}
                        H_0 - E & \ \lambda(\epsilon) \sqrt{Z(\epsilon)} \phi^{(-)}_\epsilon(\xb)  \\
                        \sqrt{Z(\epsilon)} \lambda(\epsilon) \phi^{(+)}_\epsilon(\xb) & \ \ Z(\epsilon) \left[ H_0 - E + \mu(\epsilon) \right]\\
                      \end{bmatrix} \,.
    \end{equation}
    The cut-off principal operator can be calculated algebraically by the resolvent as in the flat case and is given by
    \begin{align}\label{phi2}
      \Phi_\epsilon(E) &= Z(\epsilon) \left\{ H_0 - E + \mu(\epsilon) -\lambda^2(\epsilon) \phi^{(+)}_\epsilon(\xb) \frac{1}{H_0 - E} \phi^{(-)}_\epsilon(\xb) \right\}\nonumber\\
      &= Z(\epsilon) \Bigg\{ H_0 - E + \mu(\epsilon) -\lambda^2(\epsilon) \ix \my K_{\epsilon/2}(\xb,x) K_{\epsilon/2}(\xb,y) \nonumber \\
      & \qquad \times \ijo \mko \pjx \psky \aj \frac{1}{H_0 -E} \adk \Bigg\} \,.
    \end{align}
    Henceforth, the same game is played in order to renormalize the theory. First of all, one should normal-order this object by letting the creation operator stand on the right and the annihilation operator stand on the left in the fourth term in Eq.~(\ref{phi2}). If the following operator equalities are used,
    \begin{align}
      \frac{1}{H_0 - E} \adk &= \adk \frac{1}{H_0 - E + \ok} \,, \\
      \aj \frac{1}{H_0 - E + \ok} &= \frac{1}{H_0 - E + \ok + \oj} \aj \,,
    \end{align}
    then the principal operator becomes
    \begin{align}
      \Phi_\epsilon(E) &= Z(\epsilon) \Bigg\{ H_0 - E + \mu(\epsilon) -\lambda^2(\epsilon) \ix \my K_{\epsilon/2}(\xb,x) K_{\epsilon/2}(\xb,y)  \Bigg[ \inj \pjx \phi^\ast_j(y) \frac{1}{2 \oj} \frac{1}{H_0 - E - \oj} \nonumber \\
      & \qquad + \ijo \mko \pjx \psky \adk \frac{1}{H_0 - E + \ok + \oj} \aj \Bigg] \Bigg\} \,.
    \end{align}
    We will again use a Feynman parametrization and do an exponentiation. After that we compute the Feynman integral as
    \begin{align}
      \frac{1}{\oj} \frac{1}{H_0 - E + \oj} &= \int^1_0 d \zeta \, \frac{1}{\left[ (H_0 - E) \zeta + \oj \right]^2} \nonumber \\
      &= \int^1_0 d \zeta \, \int^\infty_0 ds \, s e^{-s \oj} e^{-s (H_0 - E) \zeta} \nonumber \\
      &= \int^\infty_0 ds \,  e^{-s \oj} \frac{1}{H_0 - E} \left[1 - e^{-s (H_0 - E)}\right] \,.
    \end{align}
    By means of the subordination identity, $\oj$ can be turned into $\osj$ which allows us to convert $e^{-s \osj}$ into a heat kernel via sandwiching it with the eigenfunctions of the operator $L$:
    \begin{equation}
      K_u(y,x) = \int d\mu(j) \, \pjx \phi^\ast_j(y) e^{-u \osj} \,.
    \end{equation}
    The reproducing identity also allows us to combine the convoluted heat kernels as
    \begin{equation}
      K_{u + \epsilon}(\xb,\xb) = \ix \my K_{\epsilon/2}(\xb,x) K_u(x,y) K_{\epsilon/2}(y,\xb) \,.
    \end{equation}
    If all of them are taken into account, we reach the following form of the principal operator,
    \begin{align}
      \Phi_\epsilon(E) &= Z(\epsilon) \Bigg\{ H_0 - E + \mu(\epsilon) - \lambda^2(\epsilon) \ix \my K_{\epsilon/2}(\xb,x) K_{\epsilon/2}(\xb,y) \nonumber \\
      & \qquad \times \ijo \mko \pjx \psky \adk \frac{1}{H_0 - E + \ok + \oj} \aj \nonumber \\
      & \qquad - \frac{\lambda^2(\epsilon)}{4\sqrt{\pi}} \int^\infty_0 du \, \int^\infty_0 ds \, s e^{-s^2/4} K_{u+\epsilon} ( \xb , \xb ) \frac{\left[ 1 - e^{- s \sqrt{u} (H_0 - E)} \right]}{\sqrt{u} ( H_0 -E )} \Bigg\} \,.
    \end{align}
    We can also exponentiate the fraction in the fourth term:
    \begin{equation}\label{ex}
      \frac{1}{H_0 - E + \ok + \oj} = \int^\infty_0 ds \, e^{-s \ok} e^{-s \oj} e^{-s (H_0 - E)} \,,
    \end{equation}
    Moreover both $\aj$ and $\adj$ can be given by the field itself by an inverse transform:
    \begin{align}\label{it}
      \aj &= \sqrt{2\oj} \int \mz \phi^\ast_j(z) \phi^{(+)}(z) \,, \\
      \adk &= \sqrt{2\ok} \int \mz \phi_k(z) \phi^{(-)}(z) \,.
    \end{align}
    Equations~(\ref{ex}), (\ref{it}) and applying reproducing property one more time brings the principal operator to the following form
    \begin{align}
      \Phi_\epsilon(E) &= Z(\epsilon) \Bigg\{ H_0 - E + \mu(\epsilon) - \frac{\lambda^2(\epsilon)}{4\sqrt{\pi}} \int^\infty_0 du \, \int^\infty_0 ds \, s e^{-s^2/4} K_{u+\epsilon} ( \xb , \xb ) \frac{\left[ 1 - e^{- s \sqrt{u} (H_0 - E)} \right]}{\sqrt{u} ( H_0 -E )} \nonumber \\
      & \qquad -\frac{\lambda^2(\epsilon)}{4 \pi}  \ix \my \int^\infty_0 ds \, s^2 \int^\infty_0 du_1 \, \frac{e^{-s^2/4u_1}}{u^{3/2}_1} \int^\infty_0 du_2 \, \frac{e^{-s^2/4u_2}}{u^{3/2}_2} \nonumber\\
      & \qquad \times K_{\epsilon/2 + u_1}(\xb,y) K_{\epsilon/2 + u_2}(\xb,x) \phi^{(-)}(y) e^{-s (H_0 - E)} \phi^{(+)}(x) \Bigg\} \,.
    \end{align}

    We are now ready to renormalize the cut-off principal operator via redefining the cut-off dependent parameters in terms of renormalized ones and divergent parts so as to cancel the divergences emerging from the normal ordering. As in the flat case we, first, determine which powers of $u$ in the $u$-integral produce divergence in the fourth term. In order to do that one should use the short-time expansion of the heat kernel which generates some inverse powers of $u$. We combine it with the powers of $u$ coming from the expansion of the exponential in $s$. Well-known short-time expansion of the heat kernel (see \cite{rosenberg}, for example) is given by
    \begin{equation}
      K_u(x,x) \simeq \frac{1}{(4 \pi u)^{d/2}} \sum_{n=0}^\infty a_n(x) u^n \,,
    \end{equation}
    where $a_n$ are universal polynomials in the curvature tensor, its covariant derivatives and various contractions thereof. The mass term in laplacian does not affect the asymptotic expansion. It can immediately be seen that only the first term in this short-time expansion contributes to divergences when it is combined with the factors coming from the exponential. The following choices are sufficient to kill all the divergences,
    \begin{align}
      \frac{\mu(\epsilon)}{\lambda^2(\epsilon)} &= \frac{\mu_R}{\lambda^2_R} + \frac{1}{4\sqrt{\pi}} \int^\infty_0 du \, \int^\infty_0 ds \, s^2 e^{-s^2/4} K_{u+\epsilon} ( \xb , \xb ) \nonumber \\
      &= \frac{\mu_R}{\lambda^2_R} + \frac{1}{2} \int^\infty_0 du \, K_{u+\epsilon} ( \xb , \xb ) \,, \\
      \frac{1}{\lambda^2(\epsilon)} &= \frac{1}{\lambda^2_R} - \frac{1}{8\sqrt{\pi}} \int^\infty_0 du \, \sqrt{u} \int^\infty_0 ds \, s^3 e^{-s^2/4} K_{u+\epsilon} ( \xb , \xb ) \nonumber \\
      &= \frac{1}{\lambda^2_R} - \frac{1}{\sqrt{\pi}} \int^\infty_0 du \, \sqrt{u} K_{u+\epsilon} ( \xb , \xb ) \,.
    \end{align}
    By taking the asymptotic limits of the integrals above, one can find out how the cut-off parameter $\epsilon$ controls the divergences and one gets
    \begin{align}
      \frac{\mu(\epsilon)}{\lambda^2(\epsilon)} &\simeq \frac{\mu_R}{\lambda^2_R} + \frac{1}{8 \pi^{3/2}} \frac{1}{\sqrt{\epsilon}} \quad \mathrm{as} \quad \epsilon \rightarrow 0^+ \,,\\
      \frac{1}{\lambda^2(\epsilon)} &\simeq \frac{1}{\lambda^2_R} + \frac{1}{8 \pi^2} \ln(\epsilon) \quad \mathrm{as} \quad \epsilon \rightarrow 0^+ \,.
    \end{align}
    It is striking that these are exactly the same results which we have found in the flat case. This equality arises from the fact that in this language short-time behavior captures the high energy behavior. Therefore, we would not expect any contribution from the geometry itself, only extreme local structure which is Euclidean determines the divergence. It is also easy to conclude that point from the short-time expansion of the heat kernel since only the first term contributes to divergences. Moreover, the first expansion coefficient $a_0$ does not contain the curvature scalar, it is just equal to 1. Yet, the geometry is very important for the spectrum of the theory. The principal operator is given in terms of the heat kernel at arbitrary times as well as its values at separate points.

    After replacing the parameters by their renormalized counterparts and successively taking the limit $\epsilon \rightarrow 0^+$, we can obtain the renormalized principal operator, which is given by
    \begin{align}\label{phiman}
      \frac{\Phi_R (E)}{\lambda_R^2} &= \frac{(H_0 - E)}{\lambda^2_R} + \frac{\mu_R}{\lambda^2_R} - \frac{1}{4\sqrt{\pi}} \int^\infty_0 du \, \int^\infty_0 ds \, s e^{-s^2/4} K_u ( \xb , \xb ) \frac{1}{\sqrt{u} ( H_0 -E )} \nonumber \\
      & \qquad \times[ 1 - s \sqrt{u} (H_0 - E) +\frac{1}{2} s^2 u (H_0 - E)^2 - e^{-s \sqrt{u} (H_0 - E) } ] \nonumber \\
      & \qquad - \frac{1}{4 \pi} \ix d^3_g y \int^\infty_0 ds \, s^2 \int^\infty_0 du_1 \, \frac{e^{-s^2/4u_1}}{u_1^{3/2}}  \int^\infty_0 du_2 \, \frac{e^{-s^2/4u_2}}{u_2^{3/2}} \nonumber \\
      & \qquad \times K_{u_1} (\xb , y) K_{u_2} (\xb , x) \phi^{(-)} (y)  e^{-s (H_0 - E)} \phi^{(+)} (x) \,.
    \end{align}
    After imposing $\Phi_R(E=\mu_p)|0\rangle \equiv 0$ and doing little algebra, one can also obtain the principal operator in terms of physical mass difference as in the flat case,
    \begin{align}
      \Phi_R ( E ) & = \left( H_0 - E + \mu_p \right)  \left\{ 1 + \frac{\lambda_R^2}{2} \int d \mu ( j ) \, \phi_j^\ast ( \overline{x} ) \phi_j ( \overline{x} ) \left[ - \frac{1}{\omega(j)^3}  +  \frac{1}{\omega ( j ) \left[H_0 - E + \omega ( j )\right]  \left[ -\mu_p + \omega (j ) \right]} \right] \right\} \nonumber \\
      & \qquad - \lambda_R^2 \ijo \mko \phi_j(\overline{x}) \phi_k^\ast(\overline{x}) \adk \frac{1}{H_0 - E + \ok + \oj} \aj \,. \label{phim}
    \end{align}

    We see that the renormalized Hamiltonian, after the same calculations done in the flat case, is given by
    \begin{equation}
      H_R = H_0 \left[ \chi_+ \otimes \chi_+^\dag + \chi_- \otimes \chi_-^\dag \right] + \mu_R \frac{1 - \sigma_3}{2} + \lambda_R \left[ \sigma_+ \phi^{(-)}(\xb) + \sigma_- \phi^{(+)}(\xb) \right] \,.
    \end{equation}
  \section{Asymptotic Limits}\label{als}
    In this section, we will study the asymptotic behavior of the operator $\Phi_R(E)$ in the limit of large number of bosons, $n \rightarrow \infty$, and the flat case is our starting point. Combining the fractions in Eq.~(\ref{phi}) by Feynman parametrization, exponentiating the resultant fractions, applying the subordination identity to $e^{-s \op}$ and taking the momentum integral, successively, brings the principal operator to the following form
    \begin{align}
      \Phi_R(E) &= \left( H_0 - E + \mu_p \right) \left\{ 1 + \frac{\lambda_R^2}{32 \pi^2} \int_0^1 d \xi \int_0^{1 - \xi} d \zeta \int_0^\infty ds \, \frac{1}{s}
      \int_0^\infty du \, \frac{e^{-1/4 u - u s^2 m^2}}{u^3} \left[ e^{- s \left[(H_0 - E) \xi - \mu_p \zeta\right]} - 1 \right]\right\} - \cdots \,.
    \end{align}
    The $u$-integral is just the integral representation of the modified Bessel function of the second kind $K_2(ms)$ multiplied with $8 m^2 s^2$. After letting $u \rightarrow u s^2$, calculating the $u$-integral and scaling $s$ as $s \rightarrow s/m$, one gets
    \begin{align}
      \Phi_R ( E ) &= \left( H_0 - E + \mu_p \right)  \left\{ 1 + \frac{\lambda_R^2}{4 \pi^2} \int_0^1 d \xi \int_0^{1 - \xi} d \zeta \int_0^\infty ds \, s K_2(s) \left[ e^{- s \left[(H_0 -E) \xi - \mu_p \zeta\right]/m} - 1 \right]\right\} - \cdots \,.
    \end{align}
    We can add and subtract $e^{-s}$ to the right hand side of this expression, which regularizes the $s$-integral and also generates a constant $-C$, which is numerically computable and approximately equal to $-2.67$. Thus, the principal operator is
    \begin{align}
      \Phi_R ( E ) &= \left( H_0 - E + \mu_p \right) \left\{ 1 - C\frac{\lambda_R^2}{4 \pi^2} + \frac{\lambda_R^2}{4 \pi^2} \int_0^1 d \xi \int_0^{1 - \xi} d \zeta \int_0^\infty ds \, s K_2(s) \left[ e^{- s \left[(H_0 - E) \xi - \mu_p \zeta\right]/m} - e^{-s} \right]\right\} - \cdots \,.
    \end{align}
    Being calculated by some mathematical software, the $s$-integral is equal to
    \begin{align}
      &\frac{5 a m \left[ 8 a^2 + \left( - 12 + \sqrt{2- 2a/m} \right) m^2 -\sqrt{2} m^{3/2} \sqrt{-a+m} \right] \pi}{20 \sqrt{1-a/m} (m - a) \left[m(a+m)\right]^{3/2}} \nonumber \\
      & \qquad  - \frac{32 (a-m)(a+m)^2 \sqrt{-a^2+m^2} {}_3 F_2 \left(1,1,5;2,\frac{7}{2};\frac{a+m}{2m} \right)}{20 \sqrt{1-a/m} (m - a) \left[m(a+m)\right]^{3/2}} \,,
    \end{align}
    where $a= (H_0 - E) \xi - \mu_p \zeta$. After converting the Hypergeometric function into an elementary function and doing some simplifications, the principal operator can be given by
    \begin{align}
      \Phi_R ( E ) &= \left( H_0 - E + \mu_p \right) \left\{ 1 - C\frac{\lambda_R^2}{4 \pi^2} + \frac{\lambda_R^2}{4 \pi^2} \int_0^1 d \xi \int_0^{1 - \xi} d \zeta \left[ \frac{5}{3} - \frac{\tilde{m}^2}{\left(\zeta - \tilde{\zeta}\right)\left(\zeta - \tilde{\zeta} -2 \tilde{m}\right)} \right. \right. \nonumber \\
      & \qquad + \left. \left. \frac{\left(\zeta - \tilde{\zeta} - \tilde{m} \right) \left[ \tilde{m}^2 + 2 \left( \zeta - \tilde{\zeta} \right) \left(  2 \tilde{m} - \zeta + \tilde{\zeta} \right) \right]}{\left( \zeta - \tilde{\zeta} \right)^{3/2} \left( 2 \tilde{m} - \zeta + \tilde{\zeta} \right)^{3/2}} \left( \pi - 2 \arcsin\sqrt{1 - \frac{\zeta}{2 \tilde{m}} + \frac{\tilde{\zeta}}{2 \tilde{m}}} \right) \right] \right\} - \cdots \,,
    \end{align}
    where $\tilde{m} \equiv \frac{m}{\mu_p}$ and $\tilde{\zeta} \equiv \frac{(H_0 - E) \xi -m }{\mu_p}$.

    Whether $\tilde{\zeta}$ is between the limits of the $\zeta$-integral or not is important for calculating this integral. Taking the integration interval of the $\xi$-integral and $H \geq m$ into account tells us that $\tilde{\zeta}$ is in the integration interval. This could cause poles since the denominators has some powers of $\zeta -\tilde{\zeta}$. If this is the case, then the integral should be defined either by a principal value prescription or by a Hadamard finite part prescription. In order to answer this question, one should expand the integrand around $\zeta=\tilde{\zeta}$ in series. For this expansion, logarithmic form of the inverse trigonometric function is more suitable:
    \begin{equation}
      - 2 \arcsin\sqrt{1 - \frac{\zeta}{2 \tilde{m}} + \frac{\tilde{\zeta}}{2 \tilde{m}}} = 2 i \ln \left( \sqrt{\frac{\zeta}{2 \tilde{m} } - \frac{\tilde{\zeta}}{2 \tilde{m}} } + i \sqrt{1 - \frac{\zeta}{2 \tilde{m} } + \frac{\tilde{\zeta}}{2 \tilde{m}} } \right) \,.
    \end{equation}
    The series expansion of the combination of the first and the second term is
    \begin{equation}
      \lim_{\zeta \rightarrow \tilde{\zeta}} \frac{5}{3} - \frac{\tilde{m}^2}{\left(\zeta - \tilde{\zeta}\right)\left(\zeta - \tilde{\zeta} -2 \tilde{m}\right)} = \frac{\tilde{m}}{2 \left( \zeta - \tilde{\zeta} \right)} + \frac{23}{12} + \frac{\zeta - \tilde{\zeta}}{8 \tilde{m}} + \mathrm{O} \left( \zeta - \tilde{\zeta} \right)^2 \,,
    \end{equation}
    and the expansion of the third term is
    \begin{align}
      \lim_{\zeta \rightarrow \tilde{\zeta}} \frac{\left(\zeta - \tilde{\zeta} - \tilde{m} \right) \left[ \tilde{m}^2 + 2 \left( \zeta - \tilde{\zeta} \right) \left(  2 \tilde{m} - \zeta + \tilde{\zeta} \right) \right]}{\left( \zeta - \tilde{\zeta} \right)^{3/2} \left( 2 \tilde{m} - \zeta + \tilde{\zeta} \right)^{3/2}} \left[ \pi + 2 i \ln \left( \sqrt{\frac{\zeta}{2 \tilde{m} } - \frac{\tilde{\zeta}}{2 \tilde{m}} } + i \sqrt{1 - \frac{\zeta}{2 \tilde{m} } + \frac{\tilde{\zeta}}{2 \tilde{m}} } \right) \right] \nonumber \\
      = - \frac{\tilde{m}}{2 \left( \zeta - \tilde{\zeta} \right)} - \frac{23}{12} + \frac{59 \left( \zeta - \tilde{\zeta} \right)}{40 \tilde{m}} + \mathrm{O} \left( \zeta - \tilde{\zeta} \right)^{3/2} \,.
    \end{align}
    It is astonishing that not only the singular parts but also the constant parts of the integrand in the expansion cancel each other and that limit of the integrand is just given by
    \begin{equation}
      \lim_{\zeta \rightarrow \tilde{\zeta}} (\mathrm{integrand}) = \frac{8}{5 \tilde{m}}  (\zeta - \tilde{\zeta}) + \mathrm{O} (\zeta - \tilde{\zeta})^{3/2} \,.
    \end{equation}
    Although $\tilde{\zeta}$ is between the integration limits, series expansion tells us that the $\zeta$ integral is an ordinary integral since the integrand does not have any poles at $\zeta=\tilde{\zeta}$. Thus, we do not need to introduce any prescription to compute this integral. After tedious calculations, the exact principal operator  can be obtained as
    \begin{align}
      \Phi_R ( E ) &= \left( H_0 - E + \mu_p \right) \left[ 1 - \left(C-\frac{7}{3}\right)\frac{\lambda_R^2}{4 \pi^2}\right] \nonumber \\
      & \qquad - 2 \frac{\lambda_R^2}{4 \pi^2} \sqrt{\left( H_0 - E - m \right) \left( H_0 - E + m \right)} \ln \left( \sqrt{\frac{H_0 - E - m}{2 m}} + \sqrt{\frac{H_0 - E + m}{2 m}} \right) \nonumber \\
      & \qquad - 2 \frac{\lambda_R^2}{4 \pi^2} \sqrt{\left( m - \mu_p \right) \left( m + \mu_p \right)} \arccos\sqrt{\frac{m - \mu_p}{2 m}} \nonumber \\
      & \qquad - \lambda_R^2 \ip \iq  \frac{\adq}{\sqrt{2 \oq}} \frac{1}{H_0 - E + \oq + \op} \frac{\app}{\sqrt{2 \op}} \,.
    \end{align}

    In the flat case, the asymptotic behavior of the principal operator in the limit of large number of bosons, that is $H_0 \geq nm \gg m > \mu_p$, is, then, given by
    \begin{align}
      \Phi_R (E) &\simeq H_0 \left[ 1 -  \left( C - \frac{7}{3} + \ln2 \right) \frac{\lambda_R^2}{4 \pi} \right] - \frac{\lambda_R^2}{4 \pi} H_0 \ln \left( \frac{H_0}{m} \right) \nonumber \\
      & \qquad - (\text{the normal-ordered interaction term}) \nonumber \\
      & \qquad + (\text{the lower order terms in } H_0) \,.
    \end{align}
    This asymptotic behavior has a striking feature, the interaction term is positive, multiplied by a minus sign gives a negative contribution, and the leading term of the renormalized principal operator is also negative. Whatever the leading behavior of this interaction term is, these two terms are enhancing the negative value of $\Phi_R(E)$. We can show the positivity of the interaction term in general by studying the same term in the manifold case. The interaction term in Eq.~(\ref{phiman}) can be written as
    \begin{align}
      & \frac{\lambda_R^2}{4\pi^2} \int^\infty_0 ds \, s^2 \left[ \int \my \int^\infty_0 du_1 \, \frac{e^{-s^2/4u_1}}{u_1^{3/2}} K_{u_1} (\xb , y) \phi^{(+)} (y) \right]^\dag \nonumber \\
      & \qquad \times e^{-s (H_0 - E)} \left[ \ix \int^\infty_0 du_2 \, \frac{e^{-s^2/4u_2}}{u_2^{3/2}} K_{u_2} (\xb , x) \phi^{(+)} (x) \right] \nonumber \\
      &= \frac{\lambda_R^2}{4\pi^2} \int^\infty_0 ds \, s^2 \underbrace{A^\dag(s) e^{-s (H_0 - E)} A(s)}_{>0} \,.
    \end{align}
    Since the integrand is positive, the interaction term is positive-definite apart from the minus sign in front. This, in turn, implies the operator to have a negative-definite sign. Henceforth, the operator $\Phi_R(E)$ can not have zero eigenvalues for $E$ positive but small. For large number of particles this proves the positivity of the energy, which is extremely important for the stability of the theory.

    Secondly, we will analyze the behavior of the principal operator on a general ultra-static manifold in the same limit. Having done similar calculations, Eq.~(\ref{phim}) becomes ready to be studied in the limit $n \rightarrow \infty$.
    \begin{align}\label{phim1}
      \Phi_R(E) &= \left( H_0 - E + \mu_p \right) \left\{ 1 - C(\overline{x},m)\frac{\lambda_R^2}{4\sqrt{\pi}} + \frac{\lambda_R^2}{4\sqrt{\pi}} \int_0^1 d\xi \int_0^{1-\xi} d \zeta \int_0^\infty d s \, s^3 \right. \nonumber \\
      & \left. \qquad \times \int_0^\infty du \, \frac{e^{- u m^2 - s^2 / 4 u}}{u^{3/2}} K_{u}(\overline{x},\overline{x}) \left[ e^{-s [(H_0 - E) \xi - \mu \zeta]} - e^{- s m} \right] \right\} - \cdots \,.
    \end{align}
    Appropriate scalings of the variables in the above equation allow us to take that limit and the operator is given by
    \begin{align}
      \Phi_R(E) &= \left( H_0 - E + \mu_p \right) \left\{ 1 - C(\overline{x},m)\frac{\lambda_R^2}{4\sqrt{\pi}} + \frac{\lambda_R^2}{4\sqrt{\pi}} \int_0^1 d\xi \int_0^{1-\xi} d \zeta \int_0^\infty d s \, \frac{s^3}{(nm)^3} \right. \nonumber \\
      & \left. \qquad \times \int_0^\infty du \, \frac{e^{- u m^2/(nm)^2 - s^2 / 4 u}}{u^{3/2}} K_{u/(nm)^2}(\overline{x},\overline{x}) \left[ e^{-s [(H_0 - E) \xi - \mu_p \zeta]/nm} - e^{- s m/nm} \right] \right\} - \cdots \,. \label{phimi}
    \end{align}
    The asymptotic behavior of the heat kernel is given by
    \begin{equation}
      \lim_{n \rightarrow \infty } K_{u/(nm)^2} (\overline{x},\overline{x}) \simeq \frac{(nm)^3}{(4 \pi u)^{3/2}} \,.
    \end{equation}
    Plugging the equation above into Eq.~(\ref{phimi}) allows us to take the $u$-integral and we get
    \begin{align}
      \Phi_R (E) & \simeq \left( H_0 - E + \mu_p \right) \Bigg\{ 1 - C(\overline{x},m)\frac{\lambda_R^2}{4\sqrt{\pi}} + \frac{\lambda_R^2}{4 \pi^2} \int_0^1 d\xi \int_0^{1-\xi} d \zeta \int_0^\infty d s \, \frac{s}{n^2} K_2\left(\frac{s}{n}\right) \nonumber \\
      & \qquad \times \left[ e^{-s [(H_0 - E) \xi - \mu_p \zeta]/nm} - e^{- s/n} \right] \Bigg\} + \cdots \,.
    \end{align}
    We should, now, be careful about the asymptotic expansion of the integral. Although asymptotic behavior of the function $K_2(s/n)$ can be used for $s$ small enough, we are not allowed to use it when $s$ becomes comparable with $n$ because the other multiplying factors do not decay sufficiently fast with $s$. Since the upper limit of the $s$-integral is at infinity, this is the case. However, if we rescale $s$ with $n$, this integral takes a form which is independent of $n$. Therefore, this expression becomes the same expression which we have found already in the previous case whose constant term $C$ is, basically, replaced by $\pi^{3/2} C(\overline{x},m)$. If one takes the next term in the short-time expansion of the heat kernel into account, then it can be seen that the contribution coming from that term is of the order of $1/n^2$, which is much smaller and hence neglected. Yet, there comes a contribution from the expansion of the first exponential, which results in a new constant $C'$, multiplying $H_0$. Thus, the leading behavior of the operator $\Phi_R(E)$ in the asymptotic limit $H_0 \gg m$ on a general ultra-static Riemannian manifold can be given by
    \begin{align}
      \Phi_R (E) &\simeq H_0 \left[ 1 - \frac{\lambda_R^2}{4 \pi^2} \left( \pi^{3/2} C(\overline{x},m) + C' - \frac{7}{3} + \ln2\right) \right] - \frac{\lambda_R^2}{4 \pi^2} H_0 \ln \left( \frac{H_0}{m} \right) \nonumber \\
      & \qquad - (\text{the normal-ordered interaction term}) \nonumber \\
      & \qquad + (\text{the lower order terms in } H_0) \,.
    \end{align}
    At this stage, we are unable to give precise asymptotic analysis of the normal-ordered interaction term, which requires a delicate study.
    We would like to call readers' attention to the fact that the same remarks, which have been done in the flat case, are also valid for the relativistic Lee model defined on a general ultra-static Riemannian manifold.

    At last, the manifold defined as $\mathcal{M} = \mathbb{R} \times \mathbb{H}^3$ will be considered as an example. $\mathbb{H}^3$ is, here, a three dimensional hyperbolic space. The reason why we study this manifold is based on the fact that its heat kernel is one of the simplest and explicitly known heat kernels.

    The heat kernel of the hyperbolic space $\mathbb{H}^n$, found in \cite{grigoryan}, and the diagonal heat kernel of $\mathbb{H}^3$ takes the form
    \begin{align}
      K_u(\bar{x},\bar{x}) &= \frac{1}{(4 \pi u)^{3/2}} \lim_{\bar{y} \rightarrow \bar{x}} \frac{\rho(\bar{x},\bar{y})}{\sinh \rho(\bar{x},\bar{y})} e^{-a^2 u - \rho(\bar{x},\bar{y})^2/4u} \nonumber \\
      &= \frac{e^{-a^2 u}}{(4 \pi u)^{3/2}} \,,
    \end{align}
    where $\rho(\overline{x},\overline{y})=\mathrm{dist}(\overline{x},\overline{y})$ is the geodesic distance on $\mathbb{H}^3$ and $-a^2$ is the constant sectional curvature. Having used the diagonal heat kernel in Eq.~(\ref{phim1}), the following operator could be obtained,
    \begin{align}
      \Phi_R (E) &= \left( H_0 - E + \mu_p \right) \left\{ 1 + \frac{\lambda_R^2}{32 \pi^2} \int_0^1 d \xi \int_0^{1-\xi} d\zeta \int_0^\infty ds \, s^3  \int_0^\infty du \, \frac{e^{-u (m^2 + a^2)}}{u^3} e^{-s^2/4u} \left[ e^{-s ( H \xi - \mu \zeta )} -1 \right] \right\} + \cdots  \,.
    \end{align}
    This is the same result, which was found already in the flat case with $m^2$ replaced by $m^2+a^2$. Thus, the space $\mathbb{H}^3$ modifies the mass term of the exact principal operator only. After this slight modification, the exact principal operator takes the form
    \begin{align}
      \Phi_R ( E ) &= \left( H_0 - E + \mu_p \right) \left[ 1 - \left(C-\frac{7}{3}\right)\frac{\lambda_R^2}{4 \pi^2}\right] - 2 \frac{\lambda_R^2}{4 \pi^2} \sqrt{\left( H_0 - E - \sqrt{m^2+a^2} \right) \left( H_0 - E + \sqrt{m^2+a^2} \right)} \nonumber \\ & \qquad \times \ln \left( \sqrt{\frac{H_0 - E - \sqrt{m^2+a^2}}{2 \sqrt{m^2+a^2}}} + \sqrt{\frac{H_0 - E + \sqrt{m^2+a^2}}{2 \sqrt{m^2+a^2}}} \right) \nonumber \\
      & \qquad - 2 \frac{\lambda_R^2}{4 \pi^2} \sqrt{\left( \sqrt{m^2+a^2} - \mu_p \right) \left( \sqrt{m^2+a^2} + \mu_p \right)} \arccos\sqrt{\frac{\sqrt{m^2+a^2} - \mu_p}{2 \sqrt{m^2+a^2}}} \nonumber \\
      & \qquad - (\text{the normal-ordered interaction term}) \,.
    \end{align}
    It is easy to see the asymptotic behavior of the operator $\Phi_R(E)$ a simple modification is sufficient to calculate it. Hence, one get the following,
    \begin{align}
      \Phi_R (E) &\simeq H_0 \left[ 1 + \frac{\lambda_R^2}{4 \pi^2} \left( \frac{7}{3} - \ln2 - C \right) \right] - \frac{\lambda_R^2}{4 \pi^2} H_0 \ln \left( \frac{H_0}{\sqrt{m^2+a^2}} \right) \nonumber \\
      & \qquad - (\text{the normal-ordered interaction term}) \nonumber \\
      & \qquad + (\text{the lower order terms in } H_0) \,,
    \end{align}
    and the normal-ordered interaction term, of course, changes drastically (see Eq.~(\ref{phiman})).
  \section{Lee model on $2+1$ dimensional Riemannian manifolds}\label{2p1}
    In this section, we  make a digression to an analysis of the two dimensional version of the Lee model. Our purpose here is two fold, we first would like to show that the two dimensional model is much simpler, which only requires a mass renormalization and secondly we would like to illustrate the power of this approach by obtaining an explicit bound on the ground state energy in each sector.

    We write the model on a Riemannian manifold in the matrix form by using a heat kernel cut-off function:
    \begin{equation}
      H_\epsilon -E = \begin{bmatrix}
                        H_0 - E & \ \lambda \phi^{(-)}_\epsilon(\xb)  \\
                         \lambda \phi^{(+)}_\epsilon(\xb) & \ \ \left[ H_0 - E + \mu(\epsilon) \right]\\
                      \end{bmatrix} \,.
    \end{equation}
    The model now neither requires a coupling constant renormalization nor a wave function one. We take the resolvent in the same way as before and find the principal operator as,
    \begin{align}
      \Phi_\epsilon(E) &= \Bigg\{ H_0 - E + \mu(\epsilon) -\lambda^2 \ix \my K_{\epsilon/2}(\xb,x) K_{\epsilon/2}(\xb,y)  \left[ \inj \pjx \phi^\ast_j(y) \frac{1}{2 \oj} \frac{1}{H_0 - E - \oj} \right. \nonumber \\
      & \qquad + \left. \ijo \mko \pjx \psky \adk \frac{1}{H_0 - E + \ok + \oj} \aj \right] \Bigg\} \,.
    \end{align}
    Following the same steps in the $3+1$ dimensional case, we end up with,
    \begin{align}
      \Phi_\epsilon(E) &= \Bigg\{ H_0 - E + \mu(\epsilon) - \frac{\lambda^2}{4\sqrt{\pi}} \int^\infty_0 du \, \int^\infty_0 ds \, s e^{-s^2/4} K_{u+\epsilon} ( \xb , \xb ) \frac{\left[ 1 - e^{- s \sqrt{u} (H_0 - E)} \right]}{\sqrt{u} ( H_0 -E )} \nonumber \\
      & \qquad -\frac{\lambda^2}{4 \pi} \ix \my \int^\infty_0 ds \, s^2 \int^\infty_0 du_1 \, \frac{e^{-s^2/4u_1}}{u^{3/2}_1} \int^\infty_0 du_2 \, \frac{e^{-s^2/4u_2}}{u^{3/2}_2} \nonumber\\
      & \qquad \times K_{\epsilon/2 + u_1}(\xb,y) K_{\epsilon/2 + u_2}(\xb,x) \phi^{(-)}(y) e^{-s (H_0 - E)} \phi^{(+)}(x) \Bigg\} \,.
    \end{align}
    Using the behaviour of the heat kernel on a two dimensional Riemannian manifold,
    \begin{equation}
      K_u(x,x) \simeq \frac{1}{(4 \pi u)} \sum_{n=0}^\infty a_n(x) u^n \,,
    \end{equation}
    we see that the principal operator becomes finite if we define a mass renormalization given by
    \begin{align}
      \mu(\epsilon) &= \mu_R + \frac{\lambda^2}{4\sqrt{\pi}} \int^\infty_0 du \, \int^\infty_0 ds \, s^2 e^{-s^2/4} K_{u+\epsilon} ( \xb , \xb ) \nonumber \\
      &= \mu_R + \frac{\lambda^2}{2} \int^\infty_0 du \, K_{u+\epsilon} ( \xb , \xb )
       \,.
    \end{align}
    As a result we find the renormalized principal operator as,
    \begin{align}\label{phiman2}
      \Phi_R (E)&= (H_0 - E + \mu_R) - \frac{\lambda^2}{4\sqrt{\pi}} \int^\infty_0 du \, \int^\infty_0 ds \, s e^{-s^2/4} K_u ( \xb , \xb ) \frac{1}{\sqrt{u} ( H_0 -E )} \nonumber \\
      & \qquad \times[ 1 - s \sqrt{u} (H_0 - E) - e^{-s \sqrt{u} (H_0 - E) } ] \nonumber \\
      & \qquad - \frac{\lambda^2}{4 \pi} \ix d^3_g y \int^\infty_0 ds \, s^2 \int^\infty_0 du_1 \, \frac{e^{-s^2/4u_1}}{u_1^{3/2}}  \int^\infty_0 du_2 \, \frac{e^{-s^2/4u_2}}{u_2^{3/2}} \nonumber \\
      & \qquad \times K_{u_1} (\xb , y) K_{u_2} (\xb , x) \phi^{(-)} (y)  e^{-s (H_0 - E)} \phi^{(+)} (x) \,.
    \end{align}
    If we now impose the physical mass condition $\Phi(E=\mu_p)|0>=0$, written in the eigenfunction expansion, we end up with, \begin{align}\label{phim21}
      \Phi_R ( E ) & = \left( H_0 - E + \mu_p \right)  \left\{ 1 + \frac{\lambda^2}{2} \int d \mu ( j ) \, \phi_j^\ast ( \overline{x} ) \phi_j ( \overline{x} ) \left[ \frac{1}{\omega ( j ) \left[H_0 - E + \omega ( j )\right]  \left[ -\mu_p + \omega (j ) \right]} \right] \right\} \nonumber \\
      & \qquad - \lambda^2 \ijo \mko \phi_j(\overline{x}) \phi_k^\ast(\overline{x}) \adk \frac{1}{H_0 - E + \ok + \oj} \aj \,.
    \end{align}
    The change in the renormalized part is important, if we recall that $\mu_p<\omega(j)$ this part is actually always positive for $E<nm$ (the interesting case from the bound state spectrum point of view). Thus the interaction term now competes with these two terms. If we evaluate the answer for the flat case we see that it is given by
    \begin{align}\label{phimf}
      \Phi_R ( E ) &= ( H_0 - E + \mu_p)+ \frac{\lambda^2}{4 \pi} \ln\left[\frac{H_0-E+m}{m-\mu_p}\right] \nonumber \\
      & \qquad - \lambda^2 \int \frac{d^2p}{(2\pi)^2} \int \frac{d^2q}{(2\pi)^2}  \frac{\adq}{\sqrt{2 \oq}} \frac{1}{H_0 - E + \oq + \op} \frac{\app}{\sqrt{2 \op}} \,.
    \end{align}
    Since the flat case is sufficiently important we will give a bound on the ground state energy for all particle sectors first and discuss the general case of manifolds afterwards. Note that if we can show that the principal operator becomes positive for sufficiently small values of $E$, this means that it is invertible, hence, it cannot have a zero eigenvalue beyond that value. This give us a lower bound on the ground state energy. To accomplish this we rewrite the principal operator in the form,
    \begin{equation}
      \Phi_R(E)=\tilde K(E)-U(E) \,,
    \end{equation}
    where
    \begin{align}
      \tilde K(E)&=( H_0 - E + \mu_p)+ \frac{\lambda^2}{4 \pi} \ln\left[\frac{H_0-E+m}{m-\mu_p}\right] \nonumber \\
      U(E) & =  \lambda^2 \int \frac{d^2p}{(2\pi)^2} \int \frac{d^2q}{(2\pi)^2}  \frac{\adq}{\sqrt{2 \oq}} \frac{1}{H_0 - E + \oq + \op} \frac{\app}{\sqrt{2 \op}} \,.
    \end{align}
    Note that for real values of $E$, we can drop the logarithm and the resulting operator is smaller than $\tilde K(E)$.  Thus following Rajeev \cite{rajeev}, we write an inequality of the form,
    \begin{equation}
      \Phi_R(E)> K(E)-U(E)=K(E)^{1/2}(1-K(E)^{-1/2}U(E)K(E)^{-1/2})K(E)^{1/2}\,,
    \end{equation}
    where $K(E)=H_0+\mu_p-E$. Hence to show that the operator $\Phi_R(E)$ to be invertible, it is sufficient to impose the condition, $||\tilde U(E)||=||K(E)^{-1/2}U(E)K(E)^{-1/2}||<1$. This will impose a condition on the ground state energy. If we write this out explicitly, after commuting the square root operators with the creation and annihilation operators of the interaction term,
    \begin{align}
      \tilde U(E) & =  \lambda^2 \int \frac{d^2p}{(2\pi)^2} \int \frac{d^2q}{(2\pi)^2} \nonumber  \\
      &\ \qquad \frac{\adq}{\sqrt{2 \oq}} \frac{1}{[H_0-E+\mu_p+\oq]^{1/2}[H_0 - E + \oq + \op][H_0-E+\mu_p+\op]^{1/2}} \frac{\app}{\sqrt{2 \op}} \,.
    \end{align}
    Now we can use the inequality $H_0>(n-1)m$ in the $n$ boson sector inside the operator and replacement of it results in a bigger operator function. Call this $\chi=(n-1)m+\mu_p-E$ and for $n>1$ we find as a result,
    \begin{align}
      \tilde U(E) & \leq  \lambda^2 \int \frac{d^2p}{(2\pi)^2} \int \frac{d^2q}{(2\pi)^2} \nonumber  \\
      &\ \qquad \frac{\adq}{\sqrt{2 \oq}} \frac{1}{[\chi+\oq]^{1/2}[\chi-\mu_p + \oq + \op][\chi+\op]^{1/2}} \frac{\app}{\sqrt{2 \op}} \,.
    \end{align}
    If we now use an extension of the Cauchy-Schwartz inequality to the Fock-Space operators, we find
    \begin{align}
      ||\tilde U(E)||  \leq  \frac{1}{2}n\lambda^2 \left[\int \frac{d^2p}{(2\pi)^2} \int \frac{d^2q}{(2\pi)^2}
      \frac{1}{\oq[\chi+\oq][\chi-\mu_p + \oq + \op]^{2}[\chi+\op]\op}\right]^{1/2} \,.
    \end{align}
    We now note that $\sqrt{\mathbf{p}^2+m^2} \geq |\mathbf{p}|=p$ and   $m>\mu_p$, and replace some the terms by these lower ones and thus preserving  direction of the inequalities,
    \begin{align}
      ||\tilde U(E)||  \leq  \frac{1}{2}n\lambda^2 \left[\int \frac{pdpd\Omega_p}{(2\pi)^2} \int \frac{qdqd\Omega_q}{(2\pi)^2}
      \frac{1}{pq[\chi +q+ p]^2[\chi+q][\chi+ p]}\right]^{1/2} \,.
    \end{align}
    Let us scale the momenta by $p=\chi \bar p, q=\chi \bar q$ we find 
    \begin{align}
      ||\tilde U(E)||  \leq  \frac{n\lambda^2}{8\pi^2} \frac{1}{\chi} \left[\int_0^\infty \int_0^\infty  \frac{d\bar pd\bar q}{[1+\bar q+ \bar p]^2[1+\bar q][1+ \bar p]}\right]^{1/2} \,.
    \end{align}
    And the last integral is finite, let us call its value as $C$, we then impose the condition,
    \begin{equation}
      \frac{n\lambda^2 C}{8\pi^2 \chi}<1 \,,
    \end{equation} 
    which guaranties that the $||\tilde U(E)||<1$ This implies the rigorous inequality on the ground state energy,
    \begin{equation}
      E_{gr}(n)\geq (n-1)m+\mu_p-\frac{\lambda^2 nC}{8\pi^2}.
    \end{equation}
    If we want the energy to be positive in all sectors this in turn brings about a bound on the coupling constant. In fact, for a global stability we should have the energy to be bounded by $(n-2)m$. But the present analysis is too crude to get a bound of this form, that requires a much more delicate analysis.
    
    Next we will work out the same problem for the Riemannian manifolds, it is simpler to work on the eigenfunction expansions. We follow the same approach and estimate the leading behavior of the term resulting from the renormalization.

    The denominator of the second term in Eq. (\ref{phim21}) can be united by Feynman parametrization as,
    \begin{equation}\label{118}
      \frac{\lambda^2}{2} \inj \frac{|\phi_j(\xb)|^2}{\oj \left[ H_0 - E + \oj \right] \left[ - \mu_p + \oj \right]} = \frac{\lambda^2}{2} \int_0^1 d \xi \int_0^{1- \xi} d \zeta \inj \frac{|\phi_j(\xb)|^2}{\left[ \oj (H_0 - E) \xi - \mu_p \zeta \right]^3} \,.
    \end{equation}
    After converting the fraction into an exponential, utilizing subordination identity and the definition of the heat kernel, this term becomes,
    \begin{equation}
      \frac{\lambda^2 }{4\sqrt{\pi}} \int_0^1 d \xi \int_0^{1- \xi} d \zeta \int_0^\infty d s \, s^3 \int_0^\infty d u \frac{ e^{-s^2 / 4 u - u m^2 - s (H_0 -E) \xi + \mu_p s \zeta} }{u^{3/2}} K_u(\xb,\xb) \,.
    \end{equation}
    Let $s \rightarrow s / (n m)$ and $u \rightarrow u / (n m)^2$, we obtain,
    \begin{equation}
      \frac{\lambda^2 }{4 \sqrt{\pi} (n m)^3} \int_0^1 d \xi \int_0^{1- \xi} d \zeta \int_0^\infty d s \, s^3 \int_0^\infty d u \frac{ e^{-s^2 / 4 u - u / n^2  - s (H_0 -E) \xi / (n m) + \mu_p s \zeta / (nm)} }{u^{3/2}} K_{u/(nm)^2}(\xb,\xb) \,.
    \end{equation}
    The asymptotic behavior of the heat kernel for large $n$ is given by,
    \begin{equation}
      \lim_{n \rightarrow \infty} K_{u/(nm)^2} (\xb,\xb) \simeq \frac{(nm)^2}{4 \pi u} \,.
    \end{equation}
    After pluging this asymptotic behavior, one gets,
    \begin{equation}
      \frac{\lambda^2}{16 \pi^{3/2} n m} \int_0^1 d \xi \int_0^{1- \xi} d \zeta \int_0^\infty d s \, s^3 \int_0^\infty d u \frac{ e^{-s^2 / 4 u - u / n^2  - s (H_0 -E) \xi / (n m) + \mu_p s \zeta / (nm)} }{u^{5/2}}\,.
    \end{equation}
    Performing the integrals results in,
    \begin{equation}
      \frac{\lambda^2}{4\pi} \frac{1}{(H_0 - E + \mu_p)} \ln\left[\frac{H_0 - E + m}{m - \mu_p}\right]\,.
    \end{equation}
    Taking the overall factor $(H_0 - E + \mu_p)$ into account, we find the same answer as the one in Eq. (\ref{phimf}) and the leading contribution of the renormalization to the principal operator in the large number of particles limit results in
    \begin{align}
      \Phi_R ( E ) &\simeq ( H_0 - E + \mu_p)+ \frac{\lambda^2}{4\pi} \ln \left[\frac{H_0-E+m}{m-\mu_p}\right] \nonumber \\
      & \qquad - \lambda^2 \ijo \mko \phi_j(\overline{x}) \phi_k^\ast(\overline{x}) \adk \frac{1}{H_0 - E + \ok + \oj} \aj \nonumber \\
      & \qquad + (\text{smaller order terms})\,.
    \end{align}
    The term in Eq. (\ref{118}) is always positive, and we see that its leading term is of smaller order. Hence can be dropped out safely without affecting the inequalities,
    \begin{equation}
      \Phi_R(E)> (H_0-E+\mu_p)^{1/2} [1-\tilde U(E)] (H_0-E+\mu_p)^{1/2}.
    \end{equation}

    We will work on a noncompact manifold, for the compact manifold case the zero mode should be worked out separately. We expand the $\tilde U(E)$ in the eigenfunction basis,
    \begin{align}
      \tilde U(E) & =  \lambda^2 \int d\mu(j) d\mu(k) \nonumber  \\
      &\ \qquad \frac{a^\dagger(j)}{\sqrt{2 \omega(k)}} \frac{\phi_{j}^*(\bar x)\phi_{k}(\bar x)}{[H_0-E+\mu_p+\omega(j)]^{1/2}[H_0 - E + \omega(j) + \omega(k)][H_0-E+\mu_p+\omega(k)]^{1/2}} \frac{a(k)}{\sqrt{2 \omega(k)}} \,.
    \end{align}
    Here we have $\omega(j)=\sqrt{\sigma_j^2+m^2}$ and we introduce again $\chi=(n-1)m-E$ (we drop the $\mu_p$ for simplicity). Recall that $H_0\geq (n-1)m$, we use this inequality, and the Cauchy-Schwartz inequality to find,
    \begin{equation}
      || \tilde U(E)|| < \frac{\lambda^2 n}{2} \left[\int d\mu(j) d\mu(k)
      \frac{|\phi_{j}(\bar x)|^2|\phi_{k}(\bar x)|^2}{\omega(j)[\chi+\omega(j)][\chi + \omega(j) + \omega(k)]^2[\chi+\omega(k)]\omega(k)}\right]^{1/2} \,.
    \end{equation}
    We use the following crude  inequality,
    \begin{equation}
      (\chi+\omega(j)+\omega(k))^2> (\chi+\omega(j))(\chi+\omega(k)) \,,
    \end{equation}
    which implies the opposite inequality for the inverse,
    \begin{align}
      || \tilde U(E)||  &< \frac{\lambda^2 n}{2} \left[\int d\mu(j) d\mu(k)
      \frac{|\phi_j(\xb)|^2|\phi_k(\xb)|^2}{\omega(j)[\chi+\omega(j)]^2[\chi+\omega(k)]^2\omega(k)}\right]^{1/2} \,\nonumber\\
      &<\frac{\lambda^2 n}{2} \left[\int d\mu(j)
      \frac{|\phi_j(\xb)|^2}{\omega(j)[\chi+\omega(j)]^2}\right] \,.
    \end{align}
    We now employ a Feynmann parametrization,
    \begin{equation}
      \frac{1}{\omega(j)[\chi+\omega(j)]^2}=\int_0^1 \frac{2\zeta d\zeta}{[\omega(j)(1-\zeta)+(\chi+\omega(j))\zeta]^3} \,.
    \end{equation}
    To make contact with the heat kernel we employ an exponentiation and then use the subordination identity to get,
    \begin{equation}
      ||\tilde{U}(E)||< \frac{1}{\sqrt{\pi}} \int_0^1 d\zeta \zeta \int_0^\infty ds \, s^3
      \int_0^\infty du \left[\int d\mu(j) \frac{e^{-s^2/4u-\omega^2(j)u}}{u^{3/2}}|\phi_j(\xb)|^2\right]e^{-s\chi \zeta} \,.
    \end{equation}
    Recognizing the heat kernel as,
    \begin{equation}
      K_u(\xb, \xb)=\int d\mu(j) |\phi(j)|^2e^{-\omega^2(j) u} \,,
    \end{equation}
    we can rewrite the desired inequality as,
    \begin{equation}
      ||\tilde U(E)||< \frac{1}{\sqrt{\pi}} \int_0^1 d\zeta \zeta \int_0^\infty s^3ds
      \int_0^\infty du [ e^{-m^2u} K_u(\xb, \xb)] \frac{e^{-s^2/4u}}{u^{3/2}}e^{-s\chi \zeta} \,.
    \end{equation} 
    We note  that for Cartan-Hadamard manifolds there is a nice inequality for the heat kernel \cite{grigoryan},
    \begin{equation}
      K_u(\bar x,\bar x)\leq \frac{C_1}{u} \,,
    \end{equation}
    where $C_1$ is a positive constant related to the geometry. This in turn implies for these manifolds that
    \begin{equation}
      ||\tilde U(E)||< \frac{1}{\sqrt{\pi}} \int_0^1 d\zeta \zeta \int_0^\infty s^3ds
      \int_0^\infty du e^{-m^2u} \frac{C_1}{u}\frac{e^{-s^2/4u}}{u^{3/2}}e^{-s\chi \zeta} \,.
    \end{equation}
    If we drop  $e^{-m^2u}$ term the integral can be easily found, we scale the $u$ variable as $sv$ and find,
    \begin{equation}
      ||\tilde U(E)||< \frac{1}{\sqrt{\pi}} \int_0^1 d\zeta \zeta \int_0^\infty ds e^{-s\chi \zeta}
      \int_0^\infty dv \frac{C_1}{v^{5/2}}e^{-1/4v}=C_1 \frac{\lambda^2 n}{\chi} \,.
    \end{equation}
    If we impose the condition, $C_1 \frac{\lambda^2 n}{\chi}<1$, then we have no zeros for $\Phi_R(E)$, and this implies a bound on the ground state energy,
    \begin{equation}
      E_{gr}(n)>(n-1)m-C_1\lambda^2 n \,.
    \end{equation}
    This shows that there is a rigorous bound on the ground state energy of the $n$ particle system. Again, one expects that these bounds are weak, that is, a better physical approximation should prove a better bound. Nevertheless the bounds that we found illustrate the power of this approach clearly.
  \section{Conclusion}
    In this paper, the construction of the relativistic Lee model on static Riemannian manifolds is studied. This construction is, basically, based on introducing an operator, the so-called principal operator, and renormalizing it successively \cite{rajeev}. Moreover, it allows us to renormalize the theory nonperturbatively. This operator, which can be regarded as a kind of effective Hamiltonian of the theory, converts a divergent linear problem in the Schr\"{o}dinger picture into a highly nonlinear but a well-defined problem. Since it is found through the resolvent in the Fock space, it is valid for all particle sectors of theory. Analysis of the behavior of the principal operator in different regimes can allow us to obtain definite information about the spectrum of the theory since the zero eigenvalues of the renormalized operator implicitly determines the bound state energies. Renormalization in this construction is established in two stages. First stage is identifying the divergences in the theory, which are tamed by a cut-off at the beginning, and then curing them by redefinitions of the appropriate parameters of the model. We show that the principal operator is free of divergences when the cut-off is removed. The second stage is specifying the renormalization conditions since there remains a finite arbitrariness in the definitions of the renormalized quantities after regularization. Since the renormalized mass of the source $\mu_R$ should, intuitively, be related to the physical mass at the lowest number of particles sector, we believe that a natural choice is to impose this condition on the renormalized principal operator. So we choose $\mu_p $ as the lowest energy solution of the equation $\Phi_R(E)|0\rangle=0$ and replace $\mu_R$ by this physical parameter.

    As shown, renormalization in the manifold case is much more complicated than the one in the flat case. The ultra-violet divergence in the theory is identified through the short-time singularity of the heat kernel, the short-time expansion of the heat kernel allows us to determine how to renormalize the bare parameters. Only the first term in the short-time expansion contributes the divergences and these can be absorbed in the redefinitions of mass and coupling constant. As known, mass and coupling constant renormalizations are not sufficient to let the theory be free of divergences so a wave function renormalization is needed. To fix the wave function renormalization constant, we start with a Hamiltonian in which a different normalization of two states of the system is allowed. In that way, we do not need to change the normalizations of the spin states after renormalization. The well-defined limit of a suitable combination of the cut-off dependent principal operator, coupling constant and wave function renormalization constant dictates the form of the constant $Z(\epsilon)$. The divergence structure in the manifold case is the same as the one in the flat case. This is, actually, not a surprising result and it stems from the fact the divergence in the theory is an ultra-violet type. We also analyze the model in an oblique light-front coordinate system as a case study in Appendix. Same results are obtained, which encourages us to confirm the results found in \cite{perry}.

    There is another unconventional alternative; where we set the wave function renormalization constant $Z(\epsilon)$ to $- \lambda_R^2 / \lambda^2(\epsilon)$. This will make $Z(\epsilon)$ positive below a certain value of the cut-off $\epsilon$, hence the lower block of the Hamiltonian multiplied by a positive divergent number. It will change the off-diagonal blocks into operators multiplied by an extra $i$. To make the Hamiltonian hermitian on $\mathbb{C}^2 \otimes \mathcal{F}_\mathcal{B} (\mathcal{H})$, we should define it through the operator,
    \begin{equation}
      H_\epsilon -E = \begin{bmatrix}
                        H_0 - E & \ \sqrt{Z(\epsilon)} \lambda(\epsilon) \phi^{(-)}_\epsilon(0)  \\
                        \sqrt{Z(\epsilon)} \lambda^\ast(\epsilon) \phi^{(+)}_\epsilon(0) & \ \ Z(\epsilon) \left[ H_0 - E + \mu(\epsilon) \right] \\
                      \end{bmatrix} \,.
    \end{equation}
    It is an intresting alternative to study.

    In Section (\ref{als}), we calculate, first, the exact principal operator in the flat case, and then analyze the asymptotic behavior of it in the large number of bosons limit. The analysis shows that the renormalization process changes the leading term distinctively with respect to the free Hamiltonian and it takes the form $- H_0 \ln H_0$. This seems to change the dynamics of the model drastically. Therefore one should be very careful how to define the quantum Hamiltonian from the constructed resolvent. Another astonishing characteristic of this result is the sign of the leading term, which is negative. Since the normal ordered interaction term has also a negative-definite sign, the total operator is negative-definite. This implies that the ground state energy is positive. In \cite{kaynak} it is shown that the quantum effective action of the large-$N$ Yukawa theory also takes a similar multiplicative contribution to the kinetic term. We, therefore, believe these results call our attentions to the point that the quantum field theoretical models should be examined in much more detail at the functional level.

    In Section (\ref{2p1}), to show the power of this approach, we look at the $2+1$ dimensional model, which only requires a mass renormalization and simpler. The model seems to have no ghosts. The cut-off Hamiltonian is well-defined. The renormalized resolvet allows us to give a rigorous bound on the ground state. The existance of the quantum Hamiltonian can be proved by the methods in Ref. \cite{dimock} in $2+1$ dimensions.

    How to construct the relativistic Lee model on a general static Riemannian manifold is addressed so far in this paper. However, the present analysis does not give adequate information how the spectrum of the theory can be build up. Although na\"{i}ve scaling arguments for the normal-ordered interaction term suggest that it gives a contribution of order $n$, a scrutiny of this contribution around the vicinity of the source hints at a stronger dependence of $n$. In light of these, it is possible that the actual contribution of the interaction term is of order $n \ln n$, that is comparable to the term generated as a result of the renormalization process. The detailed analysis of the principal operator, and hence the spectrum, requires developing new approximation methods. These questions are postponed to the future works.
  \section{Acknowledgement}
    The authors would like to thank Kayhan \"{U}lker for reading the manuscript and to thank Tongu\c{c} Rador for useful discussions.
  \section{Appendix}\label{appendix}
    In this section, we will give a brief sketch of the construction of the Lee model and the calculation of the principal operator in the light-front coordinate system, and will show that the theory in this coordinate system has the same divergence structure. The following oblique coordinate system is chosen,
    \begin{equation}
      u = t + x \,,
    \end{equation}
    where $u$ is the light-front time coordinate. The infinitesimal invariant distance element, the metric tensor and its inverse are also given by
    \begin{equation}
      ds^2 = du^2 - 2 du dx - dy^2 - dz^2 \,,
    \end{equation}
    \begin{equation}
      g_{\mu \nu} = \begin{pmatrix}
                      1 & -1 & 0 & 0 \\
                      -1 & 0 & 0 & 0 \\
                      0 & 0 & -1 & 0 \\
                      0 & 0 & 0 & -1
                    \end{pmatrix}\,, \quad g^{\mu \nu} = \begin{pmatrix}
                                                           0 & -1 & 0 & 0 \\
                                                           -1 & -1 & 0 & 0 \\
                                                           0 & 0 & -1 & 0 \\
                                                           0 & 0 & 0 & -1 \\
                                                         \end{pmatrix} \,.
    \end{equation}
    The scalar product of the coordinates and the conjugate momenta is
    \begin{equation}
      p_\mu x^\mu = p_u u + p x + \pt \cdot \xt \,
    \end{equation}
    where $x$ and $\xt$ are the longitudinal and the transverse coordinates, on the other hand  $p_u,p$ and $\pt$ are the light-front energy, the longitudinal and the transverse momenta, respectively. In the equal-time formulation, the bosonic field operator is given by
    \begin{equation}
      \phi(x , \xt) = \lip \ipt \frac{1}{\sqrt{2 p}} \left [ \lap e^{- i p x - i \pt \cdot \xt} + \ladp e^{ i p x + i \pt \cdot \xt} \right] \,.
    \end{equation}
    The equal-time commutation relations both for fields and for creation and annihilation operators are, respectively, given by
    \begin{align}
      \left[ \phi(u , x , \xt) , \phi(u , y , \yt) \right] &= \frac{1}{4} \mathrm{sgn} (x - y) \delta^{(2)} (\xt - \yt) \,, \\
      \left[ \lap , \ladq \right] &= (2 \pi)^3 \delta(p - q) \delta^{(2)}(\pt - \qt) \,.
    \end{align}
    The free Hamiltonian of the bosonic sector is
    \begin{equation}
      H_0 = \lip \ipt \lop \ladp \lap \,,
    \end{equation}
    where $\lop = \frac{m^2 + p^2 + \pt^2}{2 p}$.
    The positive and the negative frequency parts of the fields evaluated at the point zero are given by
    \begin{align}
      \phi_\epsilon^{(+)}(0) &= \lip \ipt \frac{\lap}{\sqrt{2 p}} \,, \\
      \phi_\epsilon^{(-)}(0) &= \lip \ipt \frac{\ladp}{\sqrt{2 p}} \,.
    \end{align}
    After normal-ordering the creation and annihilation operators, the principal operator takes the form
    \begin{align} \label{phi3}
      \frac{\Phi_\epsilon(E)}{\lambda^2(\epsilon)} &= Z(\epsilon) \left\{ \frac{(H_0 - E)}{\lambda^2(\epsilon)} + \frac{\mu(\epsilon)}{\lambda^2(\epsilon)} - \frac{1}{2} \lip \ipt \liq \iqt \frac{1}{\sqrt{p q}}  \lap \frac{1}{H_0 - E} \ladq \right\} \nonumber \\
      &= Z(\epsilon)\left\{ \frac{(H_0 - E)}{\lambda^2(\epsilon)} + \frac{\mu(\epsilon)}{\lambda^2(\epsilon)} - \lip \ipt \frac{1}{2p} \frac{1}{H_0 - E + \lop} \right. \nonumber \\
      & \qquad - \left. \lip \ipt \liq \iqt \frac{1}{2 \sqrt{p q}} \ladq \frac{1}{H_0 - E + \loq + \lop} \lap \right\} \,.
    \end{align}
    We do not need to use any Feynman parametrizations here and only an exponentiation is enough to complete the calculations, so the momentum integral in the fourth term in Eq. (\ref{phi3}) is, just,
    \begin{align}\label{mi}
      \lip \ipt \frac{1}{2p} \frac{1}{H_0 - E + \lop} = \int_\epsilon^\infty du \, \lip \ipt e^{- 2 u (H_0 -E) p - u (m^2 + p^2 + \pt^2)} \,.
    \end{align}
    At this stage, we should be careful about the limits of the angular part of the momentum integral. Since we work in a coordinate system which covers either the future-cone or the past-cone, after the following change of variables,
    \begin{equation}
      p^2 + \pt^2 = s^2 \quad \Rightarrow \quad p = s \cos \theta \,, \quad \pt = s \sin \theta \,,
    \end{equation}
    the integration interval of the $\theta$-integral becomes $[0,\frac{\pi}{2}]$. Equation~(\ref{mi}) is, then,
    \begin{align}
      \frac{1}{(2 \pi)^3 } \int_\epsilon^\infty du \, e^{- m^2 u} \int_0^\infty ds  \, s^2 \int_0^{\pi/2} d \theta \, \sin\theta \int_0^{2 \pi} d \phi \, e^{- 2 u (H_0 -E) s \cos\theta - u s^2} = \nonumber \\
      \frac{1}{8 (2 \pi)^2 } \int_\epsilon^\infty du \, \frac{e^{- m^2 u}}{u^{3/2}} \int_0^\infty ds \, s e^{- s^2/4} \frac{1}{ \sqrt{u} (H_0 - E)} \left[  1 - e^{- s \sqrt{u} (H_0 - E)} \right] \,.
    \end{align}
    By using the exponential representation of the fractions in the fifth term in Eq. (\ref{phi3}), the principal operator is given by
    \begin{align}
      \frac{\Phi_\epsilon(E)}{\lambda^2(\epsilon)} &= Z(\epsilon) \Bigg\{ \frac{(H_0 - E)}{\lambda^2(\epsilon)} + \frac{\mu(\epsilon)}{\lambda^2(\epsilon)} - \frac{1}{32 \pi^2 } \int_\epsilon^\infty du \, \frac{e^{- m^2 u}}{u^{3/2}} \int_0^\infty ds \, s e^{- s^2/4} \frac{\left[  1 - e^{- s \sqrt{u} (H_0 - E)} \right]}{ \sqrt{u} (H_0 - E)} \nonumber \\
      & \qquad - \frac{2}{\pi} \int_0^\infty ds \, \int_0^\infty d\alpha \,\int_0^\infty d\beta \, \lip \ipt \liq \iqt \nonumber \\
      & \qquad \times e^{- q \alpha^2 - s \loq} e^{- p \beta^2 - s \lop} \ladq e^{-s (H_0 - E)} \lap \Bigg\} \,.
    \end{align}
    With the help of the redefinitions of the mass and the coupling constant below
    \begin{align}
      \frac{\mu(\epsilon)}{\lambda^2(\epsilon)} &= \frac{\mu_R}{\lambda_R^2} + \frac{1}{32 \pi^2} \int_\epsilon^\infty du \, \frac{e^{- u m^2}}{u^{3/2}} \int_0^\infty ds \, s^2 e^{-s^2/4} \,, \\
      \frac{1}{\lambda^2(\epsilon)} &= \frac{1}{\lambda_R^2} - \frac{1}{64 \pi^2} \int_\epsilon^\infty du \, \frac{e^{- u m^2}}{u} \int_0^\infty ds \, s^3 e^{-s^2/4} \,,
    \end{align}
    one can take the limit $\epsilon \rightarrow 0^+$ after dividing both sides by $Z(\epsilon)$ and hence the renormalized principal operator takes the form,
    \begin{align}
      \frac{\Phi_R(E)}{\lambda_R^2} &= \frac{(H_0 - E)}{\lambda_R^2} + \frac{\mu_R}{\lambda_R^2} - \frac{1}{32 \pi^2 } \int_0^\infty du \, \frac{e^{- m^2 u}}{u^{3/2}} \int_0^\infty ds \, s e^{- s^2/4} \frac{1}{ \sqrt{u} (H_0 - E)} \nonumber \\
      & \qquad \times \left[1 - s \sqrt{u} (H_0 - E) + \frac{1}{2} s^2 u (H_0 - E) - e^{- s \sqrt{u} (H_0 - E)} \right] \nonumber \\
      & \qquad - \frac{2}{\pi} \int_0^\infty ds \, \int_0^\infty d\alpha \,\int_0^\infty d\beta \, \lip \ipt \liq \iqt \nonumber \\
      & \qquad \times e^{- q \alpha^2 - s \loq} e^{- p \beta^2 - s \lop} \ladq e^{-s (H_0 - E)} \lap \,.
    \end{align}
    Now to see the divergence patterns, we can again calculate the bare mass and the bare coupling constant asymptotically in $\epsilon$, as a result we find the following,
    \begin{align}
      \frac{\mu(\epsilon)}{\lambda^2(\epsilon)} &\simeq \frac{\mu_R}{\lambda_R^2} + \frac{1}{8 \pi^{3/2}} \frac{1}{\sqrt{\epsilon}} \quad \mathrm{as} \quad \epsilon \rightarrow 0^+ \,, \\
      \frac{1}{\lambda^2(\epsilon)} &\simeq \frac{1}{\lambda_R^2} + \frac{1}{8 \pi^2} \ln \epsilon \quad \mathrm{as} \quad \epsilon \rightarrow 0^+ \,.
    \end{align}
    We note that the divergences are controlled by the cut-off parameters in exactly the same way as in the previous cases. We believe this is in a cord with the discussion presented by the authors in \cite{perry} about the equivalence of the covariant perturbation theory and the light-front perturbation theory. This may be seen as another verification of this equivalence at a nonperturbative level.

    The asymptotic limit of the renormalized principal operator can, of course, be analyzed in this case, as well. For the calculations to be done for this analysis repeat themselves, we will not continue further in this direction.
  
\end{document}